%% file: paper.tex
\documentclass[a4paper,11pt]{article}
\pdfoutput=1 

\usepackage{jcappub,wrapfig,verbatim,amsthm,graphicx,hyperref,enumerate,enumitem, soul, siunitx, booktabs, tabularx, multirow} 
\usepackage{jcappub,wrapfig,verbatim,amsthm,graphicx,hyperref,enumerate,enumitem, soul, siunitx, booktabs, tabularx, multirow} 
\usepackage{comment}
\usepackage{float}
\usepackage{placeins}
\input{macros_paper.tex}                     
\usepackage[utf8]{inputenc}

\newcommand{\dd}{\mathrm d}

\usepackage{xcolor}

\title{Gravitational Redshift Constraints on the Effective Theory of Interacting Dark Energy}

\author[,a]{Sveva Castello,\footnote{\label{myfootnote} Corresponding author}}

\author[{\ref{myfootnote}},b,c]{Michele Mancarella,}

\author[a]{Nastassia Grimm,}

\author[a]{Daniel Sobral-Blanco,}

\author[d,a]{Isaac Tutusaus,}

\author[a]{and Camille Bonvin}

\affiliation[a]{D\'epartement de Physique Th\'eorique and Center for Astroparticle Physics,
Universit\'e de Gen\`eve, Quai E. Ansermet 24, CH-1211 Gen\`eve 4, Switzerland}

\affiliation[b]{Dipartimento di Fisica ``G. Occhialini", Universitá degli Studi di Milano-Bicocca, Piazza della Scienza 3, 20126 Milano, Italy}

\affiliation[c]{INFN, Sezione di Milano-Bicocca, Piazza della Scienza 3, 20126 Milano, Italy}

\affiliation[d]{Institut de Recherche en Astrophysique et Plan\'etologie (IRAP), Universit\'e de Toulouse, CNRS, UPS, CNES, 14 Av.~Edouard Belin, 31400 Toulouse, France}

\emailAdd{sveva.castello@unige.ch}
\emailAdd{michele.mancarella@unimib.it}

\abstract{Upcoming galaxy surveys provide the necessary sensitivity to measure gravitational redshift, a general relativistic effect that generates a dipole in galaxy clustering data when correlating two distinct populations of galaxies. Here, we study the constraining power of gravitational redshift within the framework of the effective theory of interacting dark energy. This formalism describes linear cosmological perturbations in scalar-tensor theories of gravity with a limited number of free functions, and allows each particle species to be coupled differently to the gravitational sector. In this work, we focus on Horndeski theories with a non-minimal coupling of dark matter to the scalar degree of freedom, yielding a breaking of the weak equivalence principle for this cosmic component, a scenario that is yet untested. We show that the dipole generated by gravitational redshift significantly breaks degeneracies and tightens the constraints on the parameters of the effective theory compared to the standard redshift-space distortion analysis solely based on the even multipoles in the galaxy correlation function, with an improvement of up to~$\sim 50\%$ for populations with a galaxy bias difference equal to 1. We make the Python package \texttt{EF-TIGRE} (\textit{Effective Field Theory of Interacting dark energy with Gravitational REdshift}) developed for this work publicly available.\footnote{~\url{https://github.com/Mik3M4n/EF-TIGRE}}
}

\begin{document}
\maketitle
\flushbottom
\section{Introduction} 
\label{sec:intro}

In recent years, a considerable amount of both theoretical and observational work in cosmology has focused on testing gravity on scales larger than individual galaxies, in the quest to unveil the unknown mechanism driving the accelerated expansion of the Universe. A powerful observational tool to achieve this consists in employing maps of galaxies in redshift space, which are directly sensitive to the peculiar velocities of the galaxies. Since different theories of gravity generate different velocity fields, comparing these measurements with theoretical predictions provides an efficient way of testing models beyond General Relativity (GR). However, the difficulty of such studies lies in the huge number of theories that have been proposed, making it impractical to compare them against the data one by one. 

A powerful approach is provided by the \textit{effective field theory (EFT) of dark energy}~\cite{Gubitosi:2012hu,Bloomfield:2012ff,Gleyzes:2013ooa,Bloomfield:2013efa,Piazza:2013pua,Gleyzes:2014rba,Gergely:2014rna, Bellini:2014fua,Lagos:2016wyv}, which encompasses a large class of theories in a unified formalism, describing a rich phenomenology beyond GR.\footnote{An alternative model-independent framework, which was developed to test gravity with cosmological observations on linear scales, is the Parameterized Post-Friedmann approach, see refs.~\cite{Baker:2012zs,Ferreira:2014mja,Skordis:2015yra}.} This formalism is based on a model-independent construction at the level of the action, providing a description of  linear cosmological perturbations in scalar-tensor theories of gravity around a homogeneous and isotropic background. In particular, it covers
Horndeski models~\cite{Horndeski1974}, which
constitute the most general class of Lorentz-invariant scalar-tensor theories with second-order equations of motion (this comprises e.g.~$f(R)$ gravity~\cite{Carroll:2003wy}, quintessence~\cite{Ratra:1987rm, Wetterich:1987fm}, Brans-Dicke models~\cite{Brans:1961sx}, kinetic braiding~\cite{Deffayet:2010qz}, Galileons/generalised Galileons~\cite{Nicolis:2008in,Deffayet:2009wt,Deffayet:2011gz}). In this paper, we will focus on models of this class, where four functions of time fully describe the perturbations~\cite{Gleyzes:2013ooa,Bellini:2014fua,Gleyzes:2014rba}: the kineticity $\aK$, the braiding $\aB$, the Planck-mass run rate $\aM$ and the tensor speed excess $\aT$.\footnote{An additional function of time, $\alpha_{\rm H}$, is required to describe ``beyond Horndeski" theories~\cite{Gleyzes:2014dya, Gleyzes:2014qga}, while four further functions are necessary to describe degenerate higher-order theories (DHOST, see refs.~\cite{Langlois:2015cwa,Crisostomi:2016czh,Langlois:2017mxy}).} Observational as well as theoretical constraints have been put on these free functions, combining the Cosmic Microwave Background (CMB), large-scale structure data and gravitational-wave detections~\cite{Baker:2017hug,Creminelli:2017sry,Sakstein:2017xjx,Ezquiaga:2017ekz,Kreisch:2017uet,Planck:2018vyg,Noller:2018wyv,Noller:2020afd,Bellini:2015xja,Creminelli:2019kjy,Salvatelli:2016mgy}. However, at the moment only the parameter $\aT$ is tightly restricted to its GR value of zero by the gravitational-wave event GW170817 and its electromagnetic counterpart, GRB170817A~\cite{LIGOScientific:2017zic}. Even if this already restricts the allowed space of theories~\cite{Baker:2017hug,Creminelli:2017sry,Sakstein:2017xjx,Ezquiaga:2017ekz}, a wide range of allowed models remains, and data from next-generation surveys are necessary to put stringent constraints on the remaining free functions~\cite{Gleyzes:2015rua,Alonso:2016suf,SpurioMancini:2018apc}.\footnote{Note that other bounds exist for functions describing theories beyond Horndeski, which further motivates our choice of restricting to Horndeski theories here. In particular, imposing the absence of gravitational-wave-induced instabilities in the dark energy sector places a tight bound of $|\alpha_{\rm H}|\lesssim 10^{-20}$ \cite{Creminelli:2019kjy}. Free functions of DHOST models can be constrained by astrophysical observations \cite{Crisostomi:2017pjs,Dima:2017pwp,Saltas:2018mxc, Saltas:2022ybg,Langlois:2017dyl}. }

Moreover, the aforementioned constraints are based on the assumption of a universal coupling of the gravitational sector to the various matter species. This means that the validity of the weak equivalence principle (WEP) is preserved. However, the universality of fifth force effects, mediated by additional degrees of freedom associated to deviations from GR, is yet untested. More specifically, fifth force effects on baryons and photons are strongly constrained (see e.g.\ ref.~\cite{Wagner:2012ui,vandeBruck:2013yxa,Brax:2013nsa,Brax:2014vva,Will:2014kxa}), while those on cold dark matter (CDM) could be significant and indeed occur in various theories~\cite{PhysRevLett.64.123, PhysRevLett.67.2926, Archidiacono:2022iuu, Barros:2018efl, Desmond:2020gzn, 1992ApJ...398..407G, PhysRev.169.1014, Hui:2009kc, Clesse:2017bsw, Wands:2012vg}, leading to a breaking of the WEP. To include such theories with distinct couplings, the effective treatment can be broadened to an extension known as \textit{effective theory of interacting dark energy}. This framework has been introduced in ref.~\cite{Gleyzes:2015pma} in the context of Horndeski theories and further developed in refs.~\cite{DAmico:2016ntq,Langlois:2017mxy} to include more general classes of scalar-tensor theories. Within this formalism,  ref.~\cite{Gleyzes:2015rua} considered a WEP-breaking scenario where photons and baryons have standard couplings to the metric, whereas CDM has a non-minimal coupling described by a single additional free function $\gamma_c$. The capability of future galaxy surveys to constrain the parameters of this framework was forecasted through a Fisher matrix approach based on a combination of the galaxy power spectrum in redshift space, the tomographic weak-lensing shear power spectrum, and the correlation between the integrated Sachs-Wolfe effect and the galaxy distribution. This analysis highlighted strong degeneracies between the effects of a fifth force and the other modifications.

As shown in ref.~\cite{Castello:2022uuu} using a phenomenological approach, these degeneracies are due to the fact that the growth of structure is affected in a similar manner by modifications in the Poisson and Euler equations. Both these equations determine how matter falls into a gravitational potential and consequently how it clusters, so their respective modifications cannot be distinguished through redshift-space distortion (RSD) analyses, which only probe the growth of structure.
Gravitational lensing does not help disentangling the two effects, since it does not probe the spatial distortion of the geometry and the time distortion separately~\cite{Bonvin:2022tii}. Luckily, ref.~\cite{Castello:2022uuu} has demonstrated that a measurement of the distortion of time through the effect of gravitational redshift on cosmological scales can break the degeneracies, since it provides a way of testing both the Euler equation (by comparing the distortion of time with the galaxy velocity), and the relation between the temporal and spatial distortion (by comparing the distortion of time with gravitational lensing~\cite{Sobral-Blanco:2021cks, Sobral-Blanco:2022oel, Tutusaus:2022cab}). 

Gravitational redshift is an effect predicted by General Relativity \cite{Einstein:1916vd} resulting in a change of wavelength when photons travel across gravitational potentials. This effect has been measured in the laboratory \cite{Pound:1959PhR}, from solar observations \cite{Lopresto:1991oxy}, on astrophysical scales from stacked galaxy clusters \cite{Kim:2004spf,Wotjak:2011Nat,Kaiser:2013nas,Jimeno:2015mr,Cai:2016ors, Zhu:2019bar}, as well as in the non-linear regime of large-scale structure~\cite{Alam:2017mnras,Alam:2017cmass}.
Here, we focus on cosmological scales described within linear perturbation theory and consider the two-point correlation function of galaxies as an observable. Together with other relativistic corrections, gravitational redshift generates a dipolar modulation in the correlation function~\cite{Bonvin:2013ogt, Croft:2013taa, Bonvin:2014owa} (or similarly an imaginary part in the power spectrum~\cite{McDonald:2009ud,Yoo:2012se}), which only appears by cross-correlating two populations of galaxies with different clustering properties, for example bright and faint galaxies.
Combining the relativistic dipole with the even multipoles generated by RSD{makes it possible to isolate the contribution due to gravitational redshift~\cite{Sobral-Blanco:2021cks}. While the even multipoles of the correlation function have been successfully measured with various surveys, see e.g.~\cite{Blake_2011,Howlett:2014opa,Alam2016:1607.03155v1,eBOSS:2020yzd}, the dipole is too small to be detected by current data in the linear regime ~\cite{Gaztanaga:2015jrs}. Detailed forecasts have however shown that it will be measurable by upcoming galaxy surveys, such as the Dark Energy Spectroscopic Instrument (DESI)~\cite{Bonvin:2015kuc,Beutler:2020evf} and the Square Kilometre Array (SKA)~\cite{Bonvin:2018ckp,Saga:2021jrh}. In particular, SKA is expected to observe close to a billion galaxies up to $z = 2$, allowing for a detection of the dipole with signal-to-noise of 80 (see section \ref{sec:results_LCDM_dipole}).

In this work, we investigate the constraining power of gravitational redshift 
within the framework of the effective theory of interacting dark energy established in refs.~\cite{Gleyzes:2015pma, Gleyzes:2015rua}. With this approach, we provide forecasts on constraints on fundamental parameters that enter at the level of the Lagrangian and cover a wide range of modified gravity theories, more specifically all Horndeski theories with an additional breaking of the WEP for CDM. Using a Monte Carlo Markov Chain (MCMC) analysis, we forecast the expected constraints from a survey like SKA phase 2 (SKA2)~\cite{Bull:2015lja}. We show that, without gravitational redshift, the parameters are strongly degenerated, leading to wide contours with two branches in most 2-dimensional projections. Such degeneracies leading to highly non-Gaussian posteriors cannot be captured by a Fisher analysis and require an MCMC approach. We show that the inclusion of gravitational redshift significantly alleviates the degeneracies and tightens the constraints on all parameters, reducing the 1\,$\sigma$ error bars by up to $\sim 50$ percent when the bias difference between the two populations is assumed to be 1, which is a reasonable value for current and upcoming galaxy surveys~\cite{Gaztanaga:2015jrs}. This is truly remarkable, since gravitational redshift can be extracted from the same data as the growth rate of structure by measuring a dipole from the cross-correlation of galaxies in addition to the standard even multipoles. 

This paper is structured as follows: In section~\ref{sec:effective_theory_IDE}, we revise the effective theory of interacting dark energy developed in refs.~\cite{Gleyzes:2015pma, Gleyzes:2015rua}, while highlighting differences to the original work in our implementation of the formalism. The expressions for the observables used for our analysis, in particular the galaxy clustering dipole arising from the general relativistic time distortion, are given in section~\ref{sec:corr_functions}. In section~\ref{sec:num_implementation}, we discuss the survey specifications and our numerical implementation of the effective theory approach in the \texttt{EF-TIGRE} (\textit{Effective Field Theory of Interacting dark energy with Gravitational REdshift}) Python code. Finally, we present the results on the constraining power of gravitational redshift in section~\ref{sec:results}, and we conclude in section~\ref{sec:conclusion}. Moreover, in appendix~\ref{app:phenomenological} we state the relation between our model parameters and commonly employed, phenomenologically--motivated parameters, while appendix~\ref{app:plots} contains supplementary plots.

\section{Effective theory of interacting dark energy}\label{sec:effective_theory_IDE}

In this work, we adopt the effective theory of interacting dark energy established in refs.~\cite{Gleyzes:2015pma, Gleyzes:2015rua}. This comprehensive formalism allows to describe a wide class of modified gravity models encompassing all Horndeski theories, while allowing for different couplings of the matter sector to gravity due to interactions in the dark sector. In particular, this enables us to consider a breaking of the WEP for the unknown CDM component. We summarize our approach in the following, referring to refs.~\cite{Gleyzes:2015pma, Gleyzes:2015rua} for details and highlighting the differences with respect to the original setup.

\subsection{Gravitational and matter sector}\label{sec:gravity_matter}
As in refs.~\cite{Gleyzes:2015pma, Gleyzes:2015rua}, we assume that the gravitational sector is described by a four-dimensional metric $g_{\mu \nu}$ and a scalar field $\phi$, with an action corresponding to the class of Horndeski theories~\cite{Horndeski1974}. The action can be written in the unitary gauge, where the constant-time hypersurfaces coincide with the uniform scalar field ones. In ADM form~\cite{Arnowitt:1962hi}, the metric can be written as\footnote{We work in units where the speed of light $c$ is set to 1.} 
\begin{equation}
    \dd s^2=-N^2\dd t^2+h_{ij}\rbr{\dd x^i+N^i\dd t}\rbr{\dd x^j+N^j\dd t}\,,
\end{equation}
where $N$ is the lapse function, $N^i$ is the shift and $h_{ij}$ the 3-dimensional spatial metric. The gravitational action can then be expressed in the general form
\begin{equation}
    S_g = \int \dd^4x\,\sqrt{-g}\,L(N, K_{ij}, R_{ij}, h_{ij}, D_i;t)\,, \label{Eq:Act_Grav}
\end{equation}
with $K_{ij}$ being the extrinsic curvature tensor and $R_{ij}$ the Ricci tensor associated to the constant time hypersurfaces.

The contribution from the matter action can be generically written as 
\begin{equation}
    S_m=\sum_I^{N_S}S_I\,,\qquad S_I=\int\dd^4x\,\sqrt{-\check g^{(I)}}L_I\rbr{\check g_{\mu\nu}^{(I)},\psi_I}\,,
\end{equation}
with each of the $N_S$ particle species being minimally coupled to the gravitational sector by a generally different effective metric $\check g_{\mu\nu}^{(I)}$. In this work, we consider contributions to the total energy density arising from baryons, CDM and radiation. We assume that the coupling of CDM to the gravitational sector is different from the one of baryons and radiation, leading to a breaking of the WEP. Thus, we cannot identify a global Jordan frame, but we choose to work in the Jordan frame of baryons and radiation, defining the corresponding metric as the gravitational metric $g_{\mu\nu}$. In this frame, CDM is allowed to be non-minimally coupled to $g_{\mu\nu}$. We denote baryons, CDM and radiation by the subscripts
$b$, $c$ and $r$ respectively.

\subsection{Background}\label{sec:background}
We consider a homogeneous and isotropic background described by the Friedmann-Lema\^itre-Robertson-Walker (FLRW) metric 
\begin{equation}
\dd s^2 = -\dd t^2+a^2(t)\delta_{ij}\dd x^i\dd x^j\,,
\end{equation}
where $a(t)$ is the scale factor. In the following, a bar denotes the $0$-th order quantities and we assume spatial flatness.
The background evolution equations can be obtained by taking variations of the homogeneous action $\bar S_g+\bar S_m$.

We use the logarithm of the scale factor $\ln(a)$ as a time variable, denoting the corresponding derivatives with a prime, and write the Friedmann equations as (see eqs.~(2.12)--(2.13) of ref.~\cite{Gleyzes:2015pma})
\begin{align} \label{eq:Hubbleparam}
    H^2 &= \frac{1}{3 M^2} \left( \bar{\rho}_{ b} + 
 \bar{\rho}_{ c} + \bar{\rho}_{ r} + \bar{\rho}_{\rm DE} \right)\,, \\
    \zeta &\equiv \frac{H'}{H}= -\frac{3}{2} -\frac{1}{2M^2H^2}\left( \bar{p}_{\rm DE}  + \bar{p}_r\right)\, . \label{Eq:Def_zeta}  
\end{align}
Here, $\rho$ denotes the energy density and $ p$ the pressure, and we assume that baryons and CDM are pressureless, $p_b = p_c = 0$. $H$ is the Hubble parameter and $M$ denotes the effective Planck mass, which in this framework is a function of time. Its time evolution is determined by the function $\aM$, defined by 
\begin{equation}
    \aM \equiv \frac{\dd \ln M^2}{\dd \ln(a) }\,. \label{Eq:Def_alphaM}
\end{equation}

The equations governing the energy densities of baryons, CDM and radiation follow from the invariance of the matter action under arbitrary diffeomorphisms. These are given in eqs.~(4.2)--(4.4) of ref.~\cite{Gleyzes:2015pma} and read
\begin{align} 
        \bar{\rho}'_b &= -3 \, \bar{\rho}_b\,, \label{eq:evol_rho_b} \\
          \bar{\rho}'_c &= -3(1 - \gamma_c) \bar{\rho}_c\,, \label{eq:evol_rho_c}\\
          \bar{\rho}_r' &= - 4 \bar{\rho}_r\,.\label{eq:evol_rho_r}
\end{align}
Hence, baryons and photons evolve in a standard way, as perfect fluids with pressure $p_b=0$ and $p_{r} = \rho_{r}/3$ respectively, while the evolution of CDM is modified by its non-minimal coupling to gravity, encoded in the free function $\gamma_c$.\footnote{The function $\gamma_c$ is given by a combination of the two functions governing the conformal and disformal coupling of CDM to gravity, see eqs.~(2.24) and (2.16)--(2.17) of ref.~\cite{Gleyzes:2015rua}. As a consequence, in this framework it is not possible to distinguish between conformal and disformal couplings. These would be distinguishable only for a component with pressure, see section~5 of ref.~\cite{Gleyzes:2015pma}.} Eqs.~\eqref{eq:Hubbleparam}--\eqref{Eq:Def_zeta} define the energy density and pressure of the dark energy component. The dark energy equation of state is then given by $w_{\rm DE}\equiv \bar{p}_{\rm DE}/\bar{\rho}_{\rm DE}$, and is included as a free parameter in the model. Inserting this into eqs.~\eqref{eq:Hubbleparam}--\eqref{Eq:Def_zeta}, we see that the dark energy density evolves as

\begin{equation}
\bar{\rho}'_{\rm DE}  = -3(1 + w_{\rm DE}) \bar{\rho}_{\rm DE}-3 \gamma_c \bar{\rho}_c+\aM (\bar{\rho}_b+\bar{\rho}_c+\bar{\rho}_r+\bar{\rho}_{\rm DE})\,. \label{eq:evol_rho_DE}
\end{equation}

It is convenient to introduce the dimensionless density parameters $\Omega_X\equiv \frac{\bar{\rho}_X}{3M^2H^2}$, and rewrite the evolution equations for $\bar\rho_b$, $\bar\rho_c$ and $\bar\rho_r$ as
\begin{align}
    \Omega_b' & = - \Omega_b \, [\aM - 3 w_{\rm DE} \Omega_{\rm DE} - \Omega_r]\,, \label{Eq:Evol_Omegab}\\
    \Omega_c' & = - \Omega_c \, [\aM -3 \gamma_c - 3 w_{\rm DE} \Omega_{\rm DE} - \Omega_r]\,, \\
    \Omega_r' & = - \Omega_r [1 +\aM- 3 w_\mathrm{DE} \Omega_{\mathrm{DE}} - \Omega_r]\,. \label{Eq:Evol_Omegar}
\end{align}
In the above equations, $\Omega_{\mathrm{DE}}$ is defined via eq.~\eqref{eq:Hubbleparam} by $\Omega_{\mathrm{DE}} = 1- \Omega_{b}-\Omega_c-\Omega_{r}$.
The Hubble parameter can be calculated from the solutions of eq.~\eqref{Eq:Evol_Omegab} to~\eqref{Eq:Evol_Omegar} by writing eq.~\eqref{Eq:Def_zeta} as
\begin{equation}
    \zeta = -\frac{3}{2} \left( 1 +\frac{1}{3} \, \Omega_r +  w_{\rm E}\Omega_{\rm DE}\right)\,, \label{Eq:Exp_zeta} 
\end{equation}
and integrating this expression while fixing as initial condition the value $H_0$ at present time (see section~\ref{sec:num_implementation} for a discussion about the value of $H_0$), 
\begin{align}
H(z)=H_0\exp\left[\frac{3}{2}\int_{\ln(a(z))}^0 \hspace{-0.5cm}\mathrm d\ln a'\left(1+\frac{1}{3}\Omega_r+w_{\rm DE}\Omega_{\rm DE} \right) \right]\, . \label{eq:H_backward}
\end{align}

\subsection{Perturbations}

To compute the equations of motions for linear perturbations, the gravitational Lagrangian in eq.~\eqref{Eq:Act_Grav} must be expanded up to second order. The coefficients of the second-order expansion can be conveniently factorized in three dimensionless combinations denoted by $\aB$, $\aK$ and $\aT$ (see section~2.3 of ref.~\cite{Gleyzes:2015pma} for details). The most general quadratic action leading to second-order equations of motion can consequently be written as
\begin{equation}\label{eq:Horndeskiaction}
\begin{split} 
 S_{\rm g}^{(2)}=  \int \mathrm{d}^3x \, \mathrm{d} t \,a^3  \frac{M^2}{2}   \bigg[ & \delta K^i_j \delta K^j_i-\delta K^2 + R\, \delta N+(1+\aT) \, \delta_2 \Big(   {\sqrt{h}}R/{a^3 }\Big)    \\
 &   +  \aK H^2 \delta N^2 + 4 \aB H \delta K \delta N  \bigg]   \, .
 \end{split}
\end{equation}
Here, $\delta_2$ denotes the second-order term in a perturbative expansion. The gravity modifications entering eq.~\eqref{eq:Horndeskiaction} are thus parameterized by four time-dependent functions: the tensor speed excess $\aT$, the kineticity $\aK$, the braiding $\aB$,\footnote{Note that the parameter $\aB$ used here corresponds to the one used in refs.~\cite{Gleyzes:2015pma, Gleyzes:2015rua}, but needs to be divided by a factor $-2$ to obtain the parameter $\aB$ originally introduced in ref.~\cite{Bellini:2014fua}.} and the effective Planck mass $M$, whose evolution is encoded in the parameter $\aM$ defined in eq.~\eqref{Eq:Def_alphaM}. The parameter $\aT$ is directly related to the speed of propagation of tensor modes $c_{\rm{T}} \equiv 1 + \aT$, which was constrained to match the speed of light up to $\sim10^{-15}$ by the multi-messenger observation of {the} binary neutron star merger GW170817 in conjunction with the gamma-ray burst GRB 170817A in 2017 \cite{LIGOScientific:2017zic}. Therefore, we set $\aT = 0$ throughout our analysis.\footnote{An open possibility is that the propagation speed of GWs exhibits a frequency dependence such that the GR value is recovered in the frequency range where the LIGO-Virgo detectors operate, while deviating from GR at lower frequencies~\cite{deRham:2018red}. This effect can be constrained with LISA or multi-band GW observations~\cite{Baker:2022eiz,LISACosmologyWorkingGroup:2022wjo}. In this work, we assume a frequency-independent GW speed. } We are thus left with the three free functions  $\aM$, $\aK$ and $\aB$, which take the values $\aK = \aB = \aM= 0$ in the special case of GR.

In order to write the perturbations equations, it is convenient to adopt the Newtonian gauge for the linearly perturbed FLRW metric,
\begin{equation}
\label{eq:perturbed_metric}
    \dd s^2=-(1+2\Psi)\dd t^2+a^2(t)(1-2\Phi)\delta_{ij}\,\dd x^i\dd x^j\,.
\end{equation}
We work in the quasi-static approximation, following the approach of ref.~\cite{Sawicki:2015zya}, where a hierarchy of terms is introduced based on the assumption that time derivatives are smaller than spatial ones. With these assumptions, the action in eq.~\eqref{Eq:Act_Grav} leads to the following Euler equations for baryons and CDM (see eqs.~(4.16) and (4.18) in  ref.~\cite{Gleyzes:2015pma}, where $\Phi$ and $\Psi$ are interchanged),\footnote{The velocity potentials $v_{b,c}$ in ref.~\cite{Gleyzes:2015pma} are related to the quantities $V_{b,c}$ used here as $v_{b,c}= - a V_{b,c}/k$.}
\begin{equation}
    V_{ b}'+V_{b}-\frac{k}{\mathcal H}\Psi=0 \,, \label{Eq:Euler_b}
\end{equation}
\begin{equation}
    V_{ c}'+V_{c}\rbr{1+3\gamma_c}-\frac{k}{\mathcal H}\Psi\rbr{1+\frac{6\gamma_c}{c_s^2\alpha\mu}\rbr{\aB-\aM+3\gamma_c\omega_cb_c}}=0\,. \label{Eq:Euler_cdm}
\end{equation}
Here, $V_{b,c}$ are the velocity potentials of baryons and CDM in Fourier space, related to the velocity fields by $\mathbf{V}_{b,c} = -i\mathbf{k}/k\,V_{b,c}\,$, and the quantities $c_s^2\alpha$ and $\mu$ are defined below. The Poisson equation reads (see eq.~(4.19) in ref.~\cite{Gleyzes:2015pma})
\begin{equation}
    -\frac{k^2}{a^2}\Psi = 4\pi G\mu\,\delta_m\bar \rho_m\,, \label{Eq:Poisson}
\end{equation}
i.e.\ the Newton constant $G$ is replaced by an effective Newton constant, $G\mu$, which depends on the gravity modifications and the non-minimal coupling $\gamma_c$ (see eq.~\eqref{Eq:mu} below). Combining eq.~\eqref{Eq:Poisson} with the continuity equation and the Euler equations gives rise to second-order evolution equations for the density (see eqs.~(4.23) and~(4.24) in ref.~\cite{Gleyzes:2015pma}),
\begin{align}
    \delta_b''& =  - \rbr{2+\zeta} \delta_b +\frac{3}{2} \Omega_{ m} \left[1 + \frac{2}{c_s^2 \alpha} \, (\aB - \aM) \, (\aB - \aM + 3 \gamma_c \omega_c b_c)\right]\, , \label{Eq:pert_b} \\
    \delta_c''& = - \left(2+ \zeta +3 \gamma_c \right) \delta_c +\frac{3}{2} \Omega_{ m} \left[1 + \frac{2}{c_s^2 \alpha} \, (\aB - \aM + 3 \gamma_c ) \, (\aB - \aM + 3 \gamma_c \omega_c b_c)\right]   \label{Eq:pert_c} \, .
\end{align}
Here, $\Omega_m$ and $\delta_m$ refer to the total matter contribution,
\begin{equation}
\Omega_m = \Omega_{b}+\Omega_{c}\,,\qquad \delta_m = (\Omega_b \delta_b+\Omega_c\delta_c)/\Omega_m\,,
\end{equation}
and $\omega_c\equiv \Omega_c/\Omega_m$ and $b_c\equiv \delta_c/\delta_m$ denote the fraction of CDM at the level of the background and perturbations respectively. The Euler, evolution and Poisson equations depend on the function $\zeta$ given in eq.~\eqref{Eq:Def_zeta}, and on the function $\mu$ defined as 
\begin{equation}
    \mu \equiv 1 + \frac{2}{c_s^2 \alpha} \, (\aB - \aM) \, (\aB - \aM + 3 \gamma_c \omega_c b_c) \,. \label{Eq:mu}
\end{equation}
Moreover, the evolution equations contain the quantity $c_s^2\alpha$, involving the speed of sound of scalar perturbations $c_s$ (given by eq.~(2.39) of ref.~\cite{Gleyzes:2015pma}),
\begin{equation}\label{eq:cs2a}
    c^2_s\alpha = -2 \left[(1 + \aB) \left(\zeta - \aM + \aB \right) + \aB' +\frac{3}{2} \Omega_m + 2 \Omega_r \right].
\end{equation}

The quantity $\alpha$, defined in eq.~(2.36) of ref.~\cite{Gleyzes:2015pma}, contains a combination of the functions $\aK$, $\aB$ and $\alpha_{\rm{D},c}$, where the latter arises from a disformal coupling of CDM to the metric $g_{\mu\nu}$. However, we note that the combination $c_s^2 \alpha$ in eq.~\eqref{eq:cs2a} does not depend on $\aK$ and $\alpha_{\rm{D},c}$. This is a consequence of the quasi-static approximation and implies that the kineticity $\aK$ remains unconstrained when adopting this limit.\footnote{We note that, even when the full set of perturbations equations beyond the quasi-static limit is considered, the kineticity is hardly constrained by CMB, baryon acoustic oscillations, and RSD data, and in general it can be fixed to arbitrary values without affecting the results for the remaining parameters \cite{Bellini:2015xja}.} Finally, the absence of ghost and gradient instabilities requires $c_s^2\alpha\geq 0$, and we note that a major advantage of the EFT formalism is that these theoretical constraints can be easily incorporated in the analysis, leading to a large restriction of the parameter space~\cite{Salvatelli:2016mgy}. We will include this condition in our analysis.

In the following, we will investigate the potential of upcoming galaxy surveys to constrain the functions $\aM$, $\aB$, $\gamma_c$ and $w_\mathrm{DE}$, highlighting the fundamental role played by gravitational redshift.

\section{Galaxy clustering observables} \label{sec:corr_functions}

The theoretical framework presented in the previous section can be constrained through observations of the distribution of galaxies across the sky. Galaxy surveys provide measurements of the number of galaxies $N$ as a function of redshift $z$ and angular position $\mathbf{\hat{n}}$ in the sky. We can use these observations to construct the number count fluctuations observable,
\begin{equation} \label{eq:Delta}
\Delta\equiv\frac{N(z,\mathbf{\hat{n}})-\bar N(z)}{\bar N(z)}\, ,
\end{equation}
where $\Bar{N}(z)$ is the average number of galaxies at redshift $z$. The quantity $\Delta$ can be computed at linear order in perturbation theory~\cite{Bonvin:2011bg, Challinor:2011bk, Yoo:2009au, Jeong:2011as, Yoo:2014CQG}, yielding\footnote{For a derivation of the relativistic effects at second order, see e.g.~\cite{Yoo:2014sfa,DiDio:2014lka,Bertacca:2014dra}.}
\begin{equation}
        \Delta(z, \mathbf{\hat{n}}) = \Delta^{\mathrm{st}}(z, \mathbf{\hat{n}}) + \Delta^{\mathrm{rel}}(z, \mathbf{\hat{n}}) \,, \label{eq:clustering_tot}
\end{equation}
where
\begin{align}
    &\Delta^{\mathrm{st}}(z,\mathbf{\hat{n}}) = b(z)\,\delta_m(z, \mathbf{\hat{n}}) - \frac{1}{\HH}\,\partial_r(\mathbf{V}\cdot\mathbf{\hat{n}}), \label{Eq:clustering_std} \\
    &\Delta^{\mathrm{rel}}(z,\mathbf{\hat{n}}) = \frac{1}{\HH}\partial_r\Psi + \frac{1}{\HH}\dot{\mathbf{V}}\cdot\mathbf{\hat{n}}+\Bigg[1-\frac{\dot{\HH}}{\HH^2}-\frac{2}{r\HH}-5\,s(z)\Bigg(1-\frac{1}{r\HH}\Bigg)\Bigg]\,\mathbf{V}\cdot\mathbf{\hat{n}}\,. \label{Eq:clustering_rel}
\end{align}
Here, a dot denotes derivatives with respect to conformal time, $r$ is the comoving distance and $\mathcal{H}$ is the conformal Hubble parameter. Eqs.~\eqref{eq:clustering_tot}--\eqref{Eq:clustering_rel} are valid in any metric theory of gravity and only rely on the assumption that photons travel along null geodesics of the metric $g_{\mu \nu}$. Eq.~\eqref{Eq:clustering_std} contains the two dominant contributions to $\Delta$, which we refer to as \textit{standard}: the first one arises from the matter density, related to the galaxy density by the linear bias $b$, whereas the second one encodes the RSD due to the peculiar velocity field of galaxies~\cite{Kaiser:1987qv}. On top of these terms, $\Delta$ is also affected by \textit{relativistic corrections} collected in eq.~\eqref{Eq:clustering_rel}: the first term encodes gravitational redshift, which changes the apparent size of a redshift bin due to the time distortion affecting galaxies situated inside of a potential well. We then have two Doppler terms depending on the velocity $\mathbf{V}$ and its time derivative.\footnote{Other contributions appearing in the full expression given in refs.~\cite{Bonvin:2011bg, Challinor:2011bk, Yoo:2009au}, including gravitational lensing, are subdominant in the redshift range relevant for this work~\cite{Jelic-Cizmek:2020pkh, Euclid:2021rez}.} Here, we note that $s$ is the magnification bias  (see e.g.~refs.~\cite{Bonvin:2013ogt, Challinor:2011bk}) arising from magnitude cuts imposed in the sample selection. Note that here we neglect the impact of the evolution bias, whose contribution is expected to be subdominant~\cite{Bonvin:2023jjq}.\footnote{The evolution bias $f^{\rm evol}$ depends on the redshift evolution of the populations of sources. It can be directly measured from the average number of galaxies once data become available (see ref.~\cite{Maartens:2021dqy} for a forecast using SKA specifications). However, as it does not affect the terms proportional to $x_c$ encoding the breaking of the WEP, we neglect it here, since we do not expect it to impact the constraining power of the dipole. Note that in general the contribution of $f^{\rm evol}$ to the dipole is suppressed with respect to the other contributions for two reasons. First, from eq.~(8) in ref.~\cite{Bonvin:2023jjq}, we see that the contributions proportional to the magnification bias difference is boosted at low redshift over those proportional to the evolution bias by a factor $\left(1 - \frac{1}{r \HH}\right)$. Secondly, following the argument in section 4.1 of ref.~\cite{Bonvin:2023jjq}, we see that the term depending on the difference $f^{\rm evol}_{\rm B}-f^{\rm evol}_{\rm F}$ vanishes, further reducing the impact of the evolution bias.}

The standard approach to analyze galaxy clustering data is to apply summary statistics. In this work, we consider the galaxy two-point correlation function,\footnote{We choose to work with the two-point correlation function rather than its Fourier-space counterpart, the power spectrum, since this makes it easier to implement the wide-angle correction in eq.~\eqref{Eq:xi1}.} $\xi \equiv \langle \Delta( \mathbf{\hat{n}}, z) \Delta(\mathbf{\hat{n}}', z') \rangle$, which can be expanded in multipoles, 
\begin{equation}
    \xi_\ell(z,d) = \frac{2 \ell +1}{2}\,\int^{1}_{-1}\,\mathrm{d}\mu\,\xi(z, d,\mu)\,P_\ell(\mu)\,.
\end{equation}
Here, $\mu$ is the cosine angle between the line of sight and the separation between the correlated galaxies (see figure~2 in ref.~\cite{Bonvin:2014owa} for an illustration), $d$ is the comoving distance between the galaxies and $P_\ell(\mu)$ denotes the Legendre polynomial of order $\ell$. 

In the distant-observer regime, the standard contributions (density fluctuations and RSD) generate three even multipoles: a monopole ($\ell=0$), a quadrupole ($\ell=2$) and a hexadecapole ($\ell=4$), which were measured by various surveys, see e.g.\ ref.~\cite{eBOSS:2020yzd}. These are given by
\begin{align}
\xi_0(z,d)&= \left[b^2(z) +\frac{2}{3} b(z) f(z)+\frac{1}{5} f^2(z)\right]\mu_0(z, d)\,, \label{Eq:xi0} \\   
\xi_2(z,d)&=-\left[\frac {4}{3} f(z) b(z) +\frac{4}{7} f^2(z)\right]\mu_2(z, d)\,, \label{Eq:xi2}\\
\xi_4(z,d)&= \frac{8}{35} f^2(z)\mu_4(z, d)\,,\label{Eq:xi4}
\end{align}
where $f=\mathrm{d}\ln(\delta_m)/\mathrm{d}\ln(a)$ is the matter growth rate, and we have defined 
\begin{align} \label{eq:mu2}
\mu_\ell(z,d)= \int\frac{\mathrm dk\,k^2}{2\pi^2} P_{\delta\delta}(k,z)j_\ell(kd)\,. 
\end{align}
Here, $P_{\delta\delta}(k,z)$ is the matter power spectrum, which can be computed by solving eqs.~\eqref{Eq:pert_b}--\eqref{Eq:pert_c}. 

The contribution to the even multipoles arising from  the general relativistic effects in $\Delta^{\rm rel}$ is suppressed by a factor of $(\HH/k)^2$ and is therefore negligible on sub-horizon scales. On the other hand, the cross-correlation of $\Delta^{\rm rel}$ with the standard terms breaks the symmetry of the two-point function $\xi$ \cite{Bonvin:2013ogt, Croft:2013taa, Yoo:2012se}, generating a dipole moment.\footnote{The cross-correlation between standard and relativistic terms also generates an octupole, which is however not sensitive to gravitational redshift~\cite{Sobral-Blanco:2021cks} and thus does not help in breaking degeneracies in our analysis. We will therefore not consider it in the following.}  
To extract this dipole from galaxy surveys, it is necessary to cross-correlate two differently biased populations of galaxies, for example a bright (B) and a faint (F) sample. Employing the Poisson equation \eqref{Eq:Poisson} and the Euler equations \eqref{Eq:Euler_b}--\eqref{Eq:Euler_cdm}, we obtain the following expression for the dipole, 
\begin{align} 
\xi^{\rm BF}_1(z,d)= \frac{\mathcal H}{\mathcal H_0} \nu_1(d,z)\Bigg\{& 3\, x_c \, (b_\B-b_\F)\, \gamma_c \left[f - {(1+z)\left(\frac{\mathcal{H}_0}{\mathcal{H}}\right)^2} \, \frac{ 3 \, \Omega_{m,0}}{c_s^2 \alpha}(\aB - \aM + 3 \gamma_c \omega_c b_c)\right]\nonumber \\
&-3 (s_\B-s_\F)\,f^2\,\rbr{1-\frac{1}{r\mathcal H}} + (b_{\rm B}-b_{\rm F})\,f\,\rbr{\frac{2}{r\mathcal H}+\frac{\dot{\mathcal H}}{\mathcal H^2}} \nonumber \\
&+5 f\rbr{b_{\rm B} s_{\rm F}- b_{\rm F} s_{\rm B}}\rbr{1-\frac{1}{r\mathcal H}}\Bigg\} \nonumber \\
&- \frac 25(b_\B-b_\F)\,f\,\frac{d}{r}\mu_2(d,z)\,.\label{Eq:xi1}
\end{align}
Here, $x_c$ is the fraction of CDM mass in a galaxy (which in general can differ from the background CDM fraction $\omega_c$ as well as from the fraction $b_c$$\equiv\delta_{c}/\delta_m$ at the level of perturbations), and 
\begin{align}
    \nu_1(z,d)=  \int\frac{\mathrm dk\,k}{2\pi^2} \mathcal{H}_0 P_{\delta\delta}(k,z)j_\ell(kd)\,.
\end{align}
The first line in eq.~\eqref{Eq:xi1} arises from the first three terms of $\Delta^{\mathrm{rel}}$ in eq.~\eqref{Eq:clustering_rel}, which are related to the density through the modified Poisson equation~\eqref{Eq:Poisson} and the Euler equation for CDM~\eqref{Eq:Euler_cdm}. These terms vanish if $\gamma_c=0$, i.e.\ if CDM is minimally coupled to the scalar field, thus obeying the Euler equation. Their combination is weighted by a factor $x_c$ due to the fact that the WEP is only broken for CDM, but not for baryons. The remaining Doppler contributions in eq.~\eqref{Eq:clustering_rel} lead to the second and third line of eq.~\eqref{Eq:xi1}. The fourth line contains the wide-angle effect, a contamination to the dipole from RSD due to the fact that the sky is not flat~\cite{Bonvin:2013ogt}.  Even though this term is suppressed by $d/r$, it is of the same order as the contributions in the first three lines, since it is proportional to $\mu_2(d)$, which is enhanced by a factor $k/\HH$ with respect to $\nu_1(d)$. We note that the dipole in eq.~\eqref{Eq:xi1}, including the wide-angle effect, exactly vanishes for a single galaxy population. This would not be the case if we had chosen to align the line-of-sight direction along one of the correlated galaxies instead of along the median direction between the galaxies. In this case, one would have a remaining ``large-angle effect" even for a single population of galaxies, as described in refs.~\cite{Gaztanaga:2015jrs,Reimberg:2015jma}).\footnote{When considering the galaxy power spectrum instead of the correlation function, the wide-angle term generates a dipole for a single population of galaxies because of the effect of the window function~\cite{Beutler:2018vpe}.} 

The galaxy velocity $\mathbf{V}$ appearing in eqs.~\eqref{Eq:clustering_std} and~\eqref{Eq:clustering_rel} is a weighted average of the velocities of baryons and CDM \cite{Gleyzes:2015rua},
\begin{equation}
    \mathbf{V}=x_c \mathbf{V}_c+(1-x_c)\mathbf{V}_b\,.\label{eq:Vgal}
\end{equation}
The velocity potentials $V_{b,c}$ can be related to the density fields $\delta_{b,c}$ via 
\begin{equation}
\label{eq:vel_potential}
    V_{b,c} = -\frac{\HH}{k}\,f_{{b,c}}(a)\,\delta_{b,c}(a,\mathbf{k})\,,
\end{equation}
where $f_{b,c}(a)\equiv \mathrm d\ln(\delta_{b,c})/\mathrm d\ln(a)$ is the growth rate of the respective species. Eq.~\eqref{eq:vel_potential} corresponds to the linear-order continuity equation, remaining valid for all particle species in the effective theory of interacting dark energy (see eqs.~(4.15) and (4.17) in ref.~\cite{Gleyzes:2015pma}). The galaxy velocity potential then becomes 
\begin{equation}
    V = -\frac{\HH}{k}\left[ x_c\,f_c\,\delta_c+(1-x_c)\,f_b\,\delta_b \right] =- \frac{\HH}{k}f\,\delta_m\,,
\end{equation}
where the  growth rate of galaxies $f$ is related to that of baryons and CDM through  
\begin{equation}
    f=x_cf_cb_c+(1-x_c)f_bb_b\,.
\end{equation}
In the following sections, we forecast the constraints on the effective theory of interacting dark energy arising from measurements of the even multipoles of $\xi$, and we compare them with the constraints obtained when the dipole is added to the analysis.

\section{Survey specifications and numerical implementation} \label{sec:num_implementation}

The effective theory framework and the galaxy clustering observables presented in the previous sections form the foundation for our work. Here, we discuss additional preliminaries necessary for the numerical analysis with our \texttt{EF-TIGRE} Python package. We define the survey specifications and choice of time evolution for the parameters we constrain. Moreover, we specify the set of observables and the fiducial models used in our analysis, and outline further details of our numerical implementation.

\subsection{Survey specifications}
We consider a survey similar to SKA2, covering 30,000 square degrees and ranging from $z=0$ to $z=1.5$. We divide this redshift range into 15  bins of size $\Delta z=0.1$ and in each of these bins we consider the survey specifications given in Table~3 in ref.~\cite{Bull:2015lja}.\footnote{The highest five redshift bins between $z=1.5$ and $z=2.0$ stated in Table~3 of ref.~\cite{Bull:2015lja} could as well be included in our analysis, but we omit them for numerical efficiency as they do not provide a significant improvement to our constraints.} We split the sample of galaxies into two populations with the same number of galaxies and model the bias of each population with the fitting function from ref.~\cite{Bull:2015lja}, 
\begin{align}
b_{\rm P}(z)= b_{1, \rm P}\exp(z\cdot b_{2,\rm P}) \pm \Delta b\, , 
\end{align}
where $\rm P$ denotes either the bright ($\rm P=\B$) or faint ($\rm P=\F$) population. We vary the four parameters $b_{1, \B}$, $b_{1, \F}$, $b_{2, \B}$ and $b_{2, \F}$, around fiducial values $b_{1,\rm P}=0.554$ and $b_{2,\rm P}=0.783$ following ref.~\cite{Bull:2015lja}. We fix $\Delta b=1$, which is consistent with the bias difference obtained from the measurement of BOSS~\cite{Gaztanaga:2015jrs}. This quantity could in principle be different for SKA2, but once the data become available, various possibilities of splitting the galaxy sample can be explored in order to boost the bias difference~\cite{Paillas:2021oli,Beutler:2020evf}, and thus the dipole signal from eq.~\eqref{Eq:xi1}.} The magnification biases $s_{\rm B}$ and $s_{\rm F}$ will be measurable from the average number of galaxies within the SKA2 survey, 
and for the purpose of our forecast we fix their values according to appendix~B of ref.~\cite{Castello:2022uuu}.

\subsection{Choices of parameterizations and time evolution of the free functions}
\label{sec:timePar}

In addition to the galaxy and magnification bias parameters, our observables depend on four free functions of time, which are the main focus of our work: the braiding $\aB$, the running of the Planck mass $\aM$, the parameter $\gamma_c$ encoding a non-minimal coupling of CDM to the metric $g_{\mu\nu}$, and the dark energy equation of state parameter $w_{\rm DE}$. 
In order to extract constraints on these functions, it is necessary to either assume a parameterized time evolution for them, or to resort to non-parametric models, typically consisting in constraining the free functions in redshift bins~\cite{Pogosian:2021mcs}. Due to the large dimensionality of the parameters space and the strong correlations between the free functions, in this work we choose to adopt the first strategy and assume a specific time evolution. 
Hence, we assume that $w_{\rm DE}$ is constant in time, whereas $\aB$, $\aM$ and $\gamma_c$ decay proportionally to the dark energy density when going backwards in cosmic history, in line with much previous literature (see e.g.~refs.~\cite{Bellini:2015xja, Alonso:2016suf, Huang:2015srv, Gleyzes:2015rua, Castello:2022uuu}). This time parameterization is motivated by the fact that the scalar field drives the accelerated expansion of the Universe, and that its impact should therefore be negligible well before acceleration started. In practice, for each parameter $p\in\{\aB, \aM,\gamma_c\}$, we write
\begin{equation}
\label{eq:p_evolution}
    p = p_0 \, \frac{1 - \Omega_m^{\rm \Lambda CDM} - \Omega_r^{\rm \Lambda CDM}}{1 - \Omega_{m, 0}^{\rm \Lambda CDM} - \Omega_{r, 0}^{\rm \Lambda CDM}}\,, 
\end{equation}
where the superscript $\Lambda$CDM denotes quantities evaluated in $\Lambda$CDM, i.e.~explicitly $\Omega_m^{\rm \Lambda CDM} = \Omega_{m, 0}^{\rm \Lambda CDM} \left(H_0/H^{\rm \Lambda CDM} (z) \right)^2\exp{(-3 \ln{a})}$ and $\Omega_r^{\rm \Lambda CDM} = \Omega_{r, 0}^{\rm \Lambda CDM}\left(H_0/H^{\rm \Lambda CDM} (z) \right)^2 \exp{(-4 \ln{a})}$. Note that here we parameterize the time evolution using the fiducial $\Lambda$CDM evolution for simplicity. Given that the background evolution is expected to be close to $\Lambda$CDM and that the choice of time parameterization involves a degree of arbitrariness, there is no compelling reason to introduce a more intricate evolution. With these assumptions, the degrees of freedom reduce to three free parameters $\aBo$, $\aMo$ and $\gamma_{c,0}$, denoting the present-time values of the functions. The final constraints on these parameters are sensitive to our choice of time parameterization, but we will argue in section \ref{sec:results} that this does not affect the main message of our work, i.e.~that the inclusion of gravitational redshift breaks several degeneracies among the parameters. 

Our choice of time parameterization is the same as in ref.~\cite{Gleyzes:2015rua} for $\aM$ and $\aB$, but it differs concerning $\gamma_c$. In ref.~\cite{Gleyzes:2015rua}, the authors assumed a constant value for the combination $\beta_\gamma = 3\sqrt{2}\gamma_c/(c_s\alpha^{1/2})$ and provided constraints on this quantity.\footnote{This parameterization was adopted to make an analogy to the standard assumption made for coupled quintessence models~\cite{Amendola:2011ie,Planck:2015bue}. } Instead, here we assume that also $\gamma_c$ evolves according to eq.~\eqref{eq:p_evolution}, thus ensuring that all free functions entering the evolution equations for the perturbations, eqs.~\eqref{Eq:pert_b}--\eqref{Eq:pert_c}, share the same time dependence. This is important, since assuming distinct time evolutions could artificially break the degeneracies between parameters, thus enhancing the constraining power of our observables. Our choice of constraining $\gamma_c$ instead of $\beta_\gamma$ also simplifies the numerical implementation in comparison to ref.~\cite{Gleyzes:2015rua}, as the expression for $c_s^2\alpha$ in eq.~\eqref{eq:cs2a}, combined with $\zeta$ in eq.~\eqref{Eq:Def_zeta}, is directly given in terms of the free functions. To summarize, the relevant degrees of freedom are $\aBo$, $\aMo$, $\gamma_{c,0}$ and $w_{\rm DE}$, which in the following we refer to as EFT parameters.

\subsection{Observables}\label{sec:observable}

We consider as our observables all possible auto- and cross-correlations between the bright and faint galaxy samples, which we expand in multipoles. This yields a total of three monopoles and three quadrupoles (bright-bright, faint-faint and bright-faint). Since the hexadecapole, given in eq.~\eqref{Eq:xi4}, is independent of population-specific biases, no additional information is obtained by considering various samples, so we only include the hexadecapole of the total galaxy population. Thus, we have in total seven even multipoles, which are evaluated at each redshift bin. We compare the constraints obtained from these even multipoles with an analysis where we also include the dipole in the cross-correlation between the bright and faint populations, which includes the effect of gravitational redshift. Note that the dipole is strictly zero for two populations with equal galaxy and magnification bias, as can be seen from eq.~\eqref{Eq:xi1}, and therefore no bright-bright or faint-faint dipole exists. We fix the minimum separation between galaxy pairs to $d_{\rm min}=20\,\mathrm{Mpc}/h$, such that the impact of non-linearities is negligible~\cite{ Bonvin:2023jjq}. Moreover, after testing that going beyond this range does not improve the resulting constraints, we fix the maximum separation to $d_{\rm max}=160\,\mathrm{Mpc}/h$.

Our choice of time parameterization for the modifications ensures that we recover GR at high redshift, well before the cosmic acceleration started. As a consequence, the constraints from the CMB on the parameters $\Omega_m$, $\Omega_b$, $\Omega_r$ at early time as well as on the primordial parameters $A_s$ and $n_s$ remain valid. Hence, we fix the values of these parameters according to the Planck  values~\cite{Planck:2018vyg} and start solving the background equations at recombination.

On the other hand, the dark energy equation of state parameter $w_{\rm DE}$ affects the late-time evolution of the Universe and can differ from -1 in Horndeski theories. We therefore keep it as a free parameter and we cannot directly set a Planck prior on it, as the Planck constraints on this parameter were obtained assuming the validity of GR at late times. We thus introduce an indirect constraint arising from the distance to last scattering, which is inferred in a model-independent manner from the position of the peaks in the CMB angular power spectrum. More precisely, we consider the Planck measurement of the comoving angular diameter distance $D_{\rm M}$ at recombination,\footnote{The distance $D_{\rm M}$ is related to the usual angular diameter distance by $D_{\rm M}=(1+z)D_{\rm A}$.} which is given by the ratio between the sound horizon at recombination, $r_\ast$, and the angular scale of the first peak of the CMB, $\theta_\ast$. We include the quantity $D_{\rm M}$ as an independent observable in addition to the multipoles, with a central value and variance determined by Planck~\cite{Planck:2018vyg}. The expected value is obtained by integrating the inverse of the Hubble parameter from today up to the redshift $z_\ast$ of recombination, 
\begin{equation}
D_{\rm M}=\int_0^{z_\ast}\, \frac{\mathrm dz}{H(z)}\,. \label{eq:dA}
\end{equation}
Note that this implies that, unlike for the constraints obtained in ref.~\cite{Gleyzes:2015rua}, we need to include radiation in the background evolution, as it becomes relevant at high redshift. On the other hand, since the multipoles are only observed at late time, it is sufficient to solve the perturbation equations starting at $z_{\rm in}=10$, where we can neglect the impact of radiation.

Lastly, we fix the value of the Hubble constant today, $H_0$, since it can be determined in a model-independent way from local measurements of the expansion rate. In practice, we should use the best-fit value obtained from supernova and Cepheid measurements~\cite{Riess:2021jrx}, but  since local measurements are currently in tension with the value inferred from Planck data, such a choice would mean that the fiducial cosmology is not $\Lambda$CDM. In particular, if $H_0$ differs from the Planck value, the theory of gravity should also be modified to ensure that the background evolution yields a distance to last scattering compatible with Planck. We therefore consider two separate fiducial models (see section~\ref{sec:fiducials}): a $\Lambda$CDM fiducial model, based on the assumption that the $H_0$ tension is due to systematic effects and will disappear in the future, where we choose $H_0$ to be compatible with the Planck constraints; and a modified gravity fiducial model, where the values of the parameters are compatible with $D_{\rm M}$ measured from Planck and with $H_0$ measured from supernovae and Cepheids. 

\subsection{Choice of fiducial models}
\label{sec:fiducials}

\begin{figure*}
    \centering    
    \includegraphics[width=.97\textwidth]{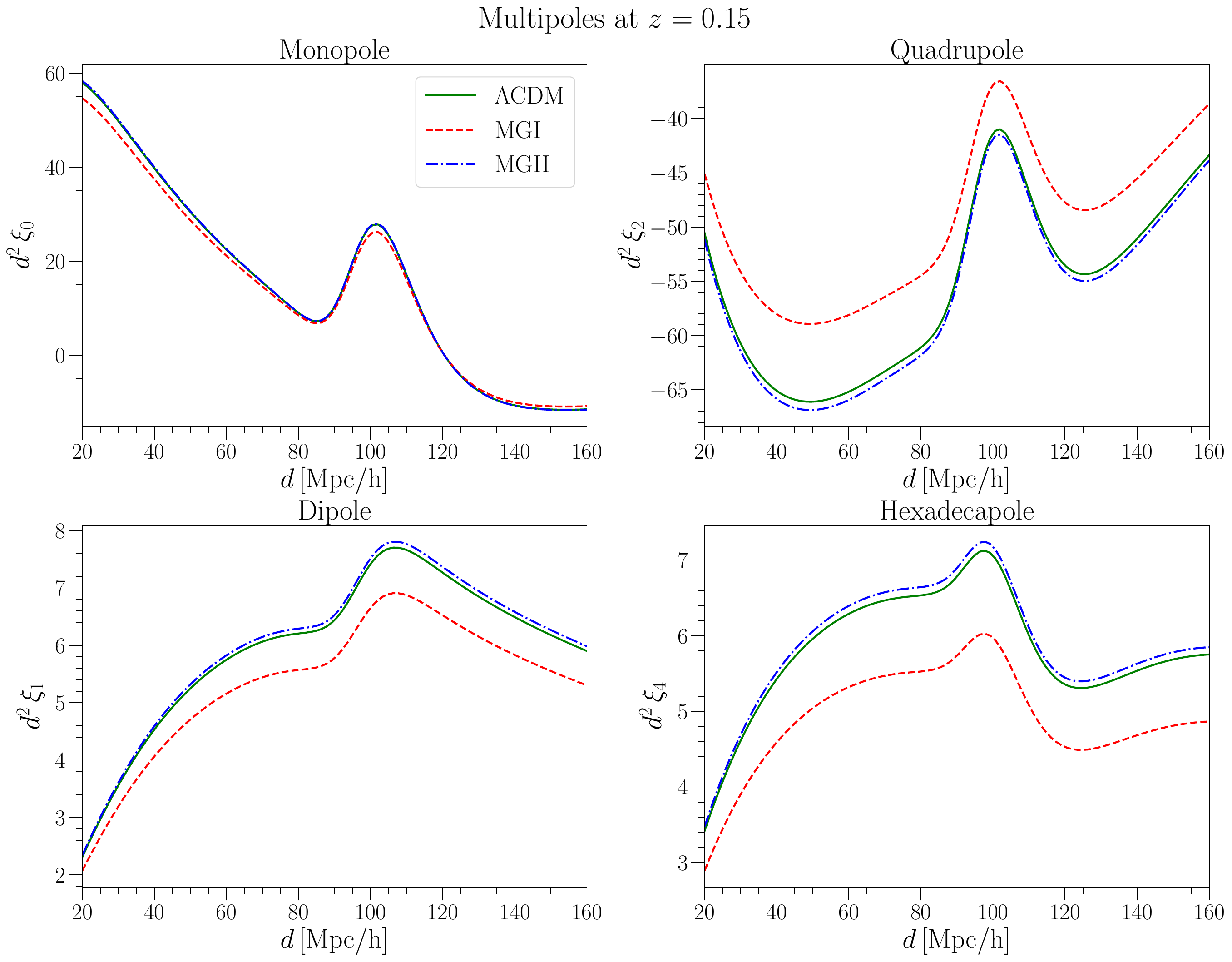}
    \caption{Multipoles of the galaxy correlation function as a function of separation, $d$, for $\Lambda$CDM (green) and the modified gravity fiducial models MGI (red) and MGII (blue), evaluated at the center of the first SKA redshift bin, $z=0.15$. }\label{fig:multipoles_plot}
\end{figure*}

We consider the three following cases:

\begin{enumerate}
\item $\Lambda$CDM fiducial with EFT parameters,
\begin{equation}
    \alpha^{\rm fid}_{\rm M,0}=\alpha^{\rm fid}_{\rm B,0}=\gamma^{\rm fid}_{c,0}=0 \, , \quad w^{\rm fid}_{\rm DE}=-1\, . \quad (\Lambda\text{CDM fiducial}) 
\end{equation}
Here we fix $H_0=67.66  \, \rm km\, s^{-1}\, Mpc^{-1}$ according to the Planck measurement~\cite{Planck:2018vyg}.

\item \emph{Modified gravity with large modifications} (MGI). In this case, we assume that $H_0=73.04  \, \rm km\, s^{-1}\, Mpc^{-1}$ is given by the local measurement of ref.~\cite{Riess:2021jrx}. In order for the distance to last scattering to match the Planck value, other parameters affecting the background evolution must be modified, i.e.\ at least one among $\alpha^{\rm fid}_{\rm M,0}$, $\gamma^{\rm fid}_{c,0}$ and $w^{\rm fid}_{\rm DE}$ should be different from the $\Lambda$CDM value. At the same time, the stability condition in eq.~\eqref{eq:cs2a} must be satisfied, which in general requires modifying at least two of these parameters. Once we adopt a modified gravity fiducial, the most general choice is that all parameters deviate from their $\Lambda$CDM values, unless some fine tuning is present. We thus choose fiducial values $\alpha^{\rm fid}_{\rm M,0}$, $\alpha^{\rm fid}_{\rm B,0}$ and  $\gamma^{\rm fid}_{c,0}$, which lie between the $1$ and $2\,\sigma$ constraints of the $\Lambda$CDM fiducial model. Then, the value of $w^{\rm fid}_{\rm DE}$ is fixed such that the distance to last scattering matches the Planck measurement when assuming $H_0=73.04  \, \rm km\, s^{-1}\, Mpc^{-1}$. In summary, we choose for the EFT parameters
\begin{equation}
    \alpha^{\rm fid}_{\rm M,0}=-0.1\, , \quad \alpha^{\rm fid}_{\rm B,0}=-0.5\, ,\quad \gamma^{\rm fid}_{c,0}=0.15 \, ,\quad  w^{\rm fid}_{\rm DE}=-1.41\, . \quad (\text{MGI fiducial}) 
\end{equation}
Note that $w_{\rm DE}<-1$ does not lead to instabilities in Horndeski theories, since the effect of the equation of state on the sound speed can be counterbalanced by other parameters to keep $c_s^2\alpha>0$ (see eq.~\eqref{eq:cslin} below).

\item \emph{Modified gravity with small modifications} (MGII). The values of the MGI fiducial model correspond to quite large deviations from $\Lambda$CDM. For comparison, we also consider a fiducial model with smaller deviations, to investigate the impact on the constraints with and without the gravitational redshift. 
We therefore include a third fiducial model defined by 
\begin{equation}
    \alpha^{\rm fid}_{\rm M,0}=-0.01\, , \quad \alpha^{\rm fid}_{\rm B,0}=-0.1\, ,\quad \gamma^{\rm fid}_{c,0}=0.01 \, ,\quad  w^{\rm fid}_{\rm DE}=-1.01\, . \quad (\text{MGII fiducial}) 
\end{equation}
In this case, $H_0$ is fixed to the Planck value, $\alpha^{\rm fid}_{\rm M,0}$, $\alpha^{\rm fid}_{\rm B,0}$ and  $\gamma^{\rm fid}_{c,0}$ are chosen arbitrarily, and $w^{\rm fid}_{\rm DE}$ is determined from the distance to last scattering. 

\end{enumerate}
We always fix the values of $A_s$ and $n_s$ and the density parameters $\Omega_m$, $\Omega_b$, $\Omega_r$ at $z_*$ according to the Planck constraints \cite{Planck:2018vyg}. We show the galaxy correlation multipoles in the three fiducial models in figure~\ref{fig:multipoles_plot}. The deviation of the even multipoles with respect to $\Lambda$CDM is due to a shift in the growth rate $f$, which is reduced in MGI and increased in MGII. The monopole is less affected by gravity modifications because it is dominated by the density contribution, which is less sensitive to deviations from GR than the velocity contribution. The dipole is additionally affected by the term due to gravitational redshift in eq.~\eqref{Eq:xi1}, which is zero in $\Lambda$CDM and positive in both MGI and MGII.

In addition to the equation of state and EFT parameters, the observables are affected by the fraction of CDM within a galaxy, $x_c$, defined in eq.~\eqref{eq:Vgal}, which is generally unknown. The most straightforward possibility would be to assume that this fraction is equal to the fraction of CDM at the background level, as was done in ref.~\cite{Gleyzes:2015rua}. However, this is not necessarily the case. Therefore, we include a free parameter $X$ in our analysis, capturing the deviation of $x_c$ from the background CDM fraction $\omega_c$. 
In particular, we assume that the fractions of mass in CDM and baryons in each galaxy (denoted as $x_c$, $x_b$ respectively) are related to the  respective background energy fractions $\omega_b$, $\omega_c$ as
\begin{align}
    x_c = X \omega_c \, , \qquad x_b = X \omega_b +1-X \, , \label{eq:xc_xb}
\end{align}
so that the condition $x_c+x_b=1$ is always satisfied. The parameter $X$ can have values $X \in [ 0, 1+\omega_b/\omega_c]$, corresponding to the extreme cases $x_c=1,\ x_b=0$ for $X=1+\omega_b/\omega_c$, and $x_c=0,\ x_b=1$ for $X=0$.\footnote{Note that, if the WEP is satisfied, then $\omega_b$, $\omega_c$ are constant, and the domain of $X$ is fixed by the initial conditions. In our case, however, baryons and CDM can evolve differently, so the upper bound for $X$ depends on the background evolution and in particular on the value of $\gamma_c$. In practice, during the MCMC analysis, at each step we reject the sample if $X$ lies outside the range $[ 0, 1+\omega_b/\omega_c]$.} The fiducial value is assumed to be $X=1$, in which case the fraction of CDM inside a galaxy corresponds to the background one.
 
\subsection{MCMC analysis} \label{sec:numerical_details}

In summary, our model is characterized by the following set of parameters:
\begin{equation}
(\aBo,~\aMo,~\gamma_{c,0},~w_{\rm DE},~X,~b_{1,\B},~b_{2,\B},~b_{1,\F},~b_{2,\B})\, . \label{Eq:params}
\end{equation}
For each fiducial model in section~\ref{sec:fiducials}, we generate mock values for the observables described above assuming zero noise. Then, we perform an MCMC analysis proceeding  as follows. First, we evaluate the background quantities $\Omega_m$, $\Omega_c$ and $\Omega_r$ by solving the system of equations given in eqs.~\eqref{Eq:Evol_Omegab}--\eqref{Eq:Evol_Omegar}. 
We then obtain $H(z)$ by integrating eq.~\eqref{eq:H_backward} with initial condition $H(z=0)=H_0$, and compute the combination $c_s^2\alpha$  from eq.~\eqref{eq:cs2a} with $\zeta$ given by eq.~\eqref{Eq:Def_zeta}. Note that the derivative $\aB'$ can be expressed using the time evolution specified in eq.~\eqref{eq:p_evolution}. If $c_s^2\alpha$ is negative, signaling the presence of an instability, the point is rejected. Otherwise, we insert the background solutions into the perturbations equations, solve the resulting system of equations in eqs.~\eqref{Eq:pert_b} and~\eqref{Eq:pert_c}, and compute the expected values of all the observables.

The likelihood for the galaxy clustering observables is given by a multivariate Gaussian centered on the fiducial values. The respective covariance matrix includes cosmic variance due to the finite survey volume as well as shot noise arising from the finite number of observed galaxies (see appendix~C of ref.~\cite{Bonvin:2018ckp} for detail). We include the non-vanishing cross-correlations between the even multipoles as well as between the bright and faint sample of galaxies. We also consider the pure cosmic variance contribution to the dipole from relativistic effects following ref.~\cite{Bonvin:2023jjq}, where the robustness of the full theoretical expression for the covariance was thoroughly tested.\footnote{In particular, ref.~\cite{Bonvin:2023jjq} performed a comparison to the simulation-based jackknife covariance in the case of DESI, showing that the theoretical covariance that we also adopt here slightly underestimates the errors due to the impact of non-linearities. This only leads to a mild difference in the detectability of the dipole in the case of DESI. For SKA2, we expect a similar impact in the low-redshift sample, but a lower impact at high redshifts where non-linearities have less importance. Note that the complete expression for the covariance should also contain the impact of the survey mask. This can be evaluated once the survey is performed.} For the angular diameter distance to the CMB, we assume a Gaussian likelihood centered on the Planck value $D_{\rm M} = r_*/\theta_* $, where $r_*$ and $\theta_*$ are taken from Planck 2018~\cite{Planck:2018vyg}. The variance is computed by propagating the error accordingly. We adopt large priors on all variables except for $X$, which is restricted to its physical range as already described in section~\ref{sec:fiducials}. 

We use the ensemble slice sampler \texttt{zeus}~\cite{Karamanis:2020zss, Karamanis:2021tsx} with 18 walkers, ensuring that the lechangeh of the chains is at least 50 times the autocorrelation time, and further excluding a burn-in phase equal to the correlation lechangeh and thinning by the autocorrelation time.

\section{Results}
\label{sec:results}

In this section, we present the constraints obtained on the EFT parameters $\aMo$, $\aBo$, $\gamma_{c,0}$ and $w_{\rm DE}$. We discuss the results for the three fiducial models introduced in section \ref{sec:fiducials} ($\Lambda$CDM and two modified gravity fiducials) and we additionally study the impact of restricting the parameter space to specific subclasses of models. We compare the analysis solely based on the even multipoles of the galaxy two-point correlation function (standard RSD analysis) with the constraints obtained when also including the dipole, which contains the contribution of gravitational redshift.

\subsection{Constraints around a \texorpdfstring{$\Lambda$}{Lg}CDM fiducial model}
\label{sec:LCDMres}

\begin{figure*}
    \centering    
 \includegraphics[width=.75\textwidth]{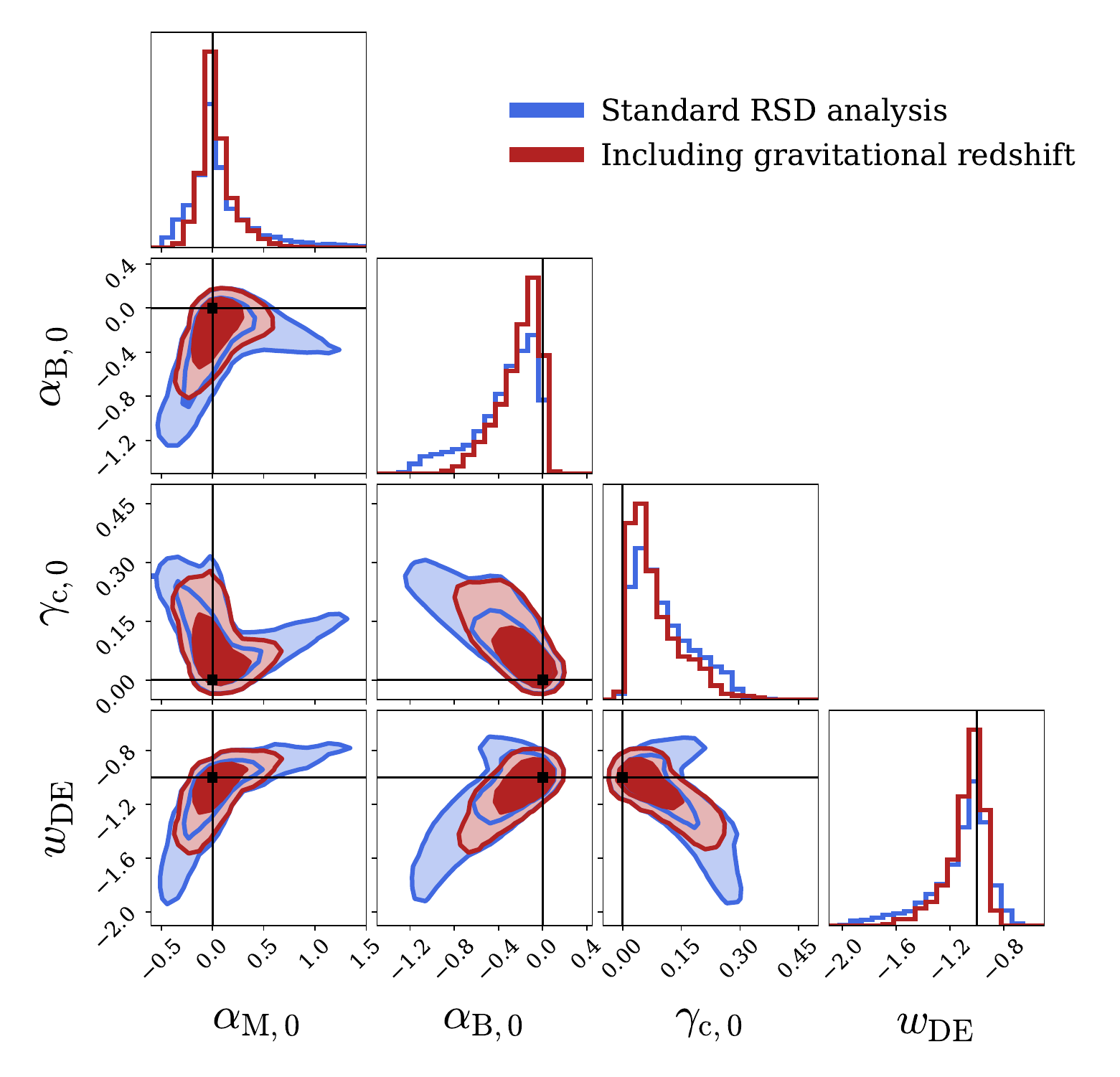}
    \caption{Constraints around a $\Lambda$CDM fiducial model. } \label{fig:LCDM_even_all}
\end{figure*}

Choosing $\Lambda$CDM as our fiducial model, we show in figure~\ref{fig:LCDM_even_all} the constraints on the EFT parameters, marginalized over the biases and the parameter $X$. The results obtained with a standard RSD analysis are presented in blue, while those that also include gravitational redshift are in red. In both cases, we also consider the angular diameter distance in our set of observables. In Table~\ref{tab:LCDMconstraints}, we report the marginal median values and credible intervals for the EFT parameters, both with and without the gravitational redshift (``Dipole" and ``No dipole" columns).

\begin{table}
\renewcommand{\arraystretch}{1.5}
  \centering
  \begin{tabular}{l|rr|ccr}
    \hline\hline
     & \multicolumn{2}{c|}{Median} & \multicolumn{3}{c}{Width of the $68\%$ credible interval}  \\
     & No dipole & Dipole  & No dipole & Dipole & Reduction \\
    \hline\hline
    $\alpha_{\rm M,0}$ & $0.01$ & $0.01$ & $[-0.18, 0.31]$ & $[-0.11, 0.17]$ & 43\% \\
    $\alpha_{\rm B,0}$ & $-0.26$ & $-0.17$ & $[-0.37, 0.19]$ & $[-0.24, 0.13]$ & 34\% \\
    $\gamma_{\rm c,0}$ & $0.09$ & $0.06$ & $[-0.05, 0.11]$ & $[-0.04, 0.09]$ & 19\% \\
    ${w}_{\rm DE}$ & $-1.06$ & $-1.05$ & $[-0.27, 0.12]$ & $[-0.16, 0.08]$ & 38\% \\
    \hline\hline
  \end{tabular} 
  \caption{\label{tab:LCDMconstraints} Median values and 68\% credible intervals for the $\Lambda$CDM fiducial model both for the standard RSD analysis (no dipole) and when adding the dipole that contains gravitational redshift. We also show the percent reduction on the credible interval obtained by including the dipole.} 
\end{table}

\subsubsection{Standard RSD analysis including even multipoles only}

When considering the even multipoles only, the constraints arise from the galaxy velocity field and the angular diameter distance to last scattering, where the latter depends on the integral of the inverse Hubble parameter $H^{-1}(z)$ from today to the redshift of decoupling. In the following, we discuss the respective constraints (blue contours in figure~\ref{fig:LCDM_even_all}) for each parameter pair. 

\begin{itemize}
\item $w_{\rm DE}-\aMo$ contour: These two parameters affect directly the evolution of $H(z)$ and consequently the angular diameter distance to the CMB. From eq.~\eqref{eq:H_backward} we see indeed that $H(z)$ is directly sensitive to $w_{\rm DE}$, and that it also depends on $\aM$ through the evolution of the radiation density $\Omega_r$, see~eq.~\eqref{Eq:Evol_Omegar}. This leads to a strong degeneracy between $\aMo$ and $w_{\rm DE}$, as can be seen in figure~\ref{fig:LCDM_even_all}. For example, an equation of state $w_{\rm DE}<-1$ leads to a dark energy pressure more negative than in $\Lambda$CDM, increasing the distance to last scattering. This can be compensated by a negative $\aM$, which makes radiation decay more slowly than in $\Lambda$CDM: the increase in radiation pressure counteracts the negative pressure of dark energy, keeping the distance in agreement with CMB constraints. On the other hand, when $w_{\rm DE}>-1$, the opposite occurs, favoring a positive $\aMo$. This leads to the two branches in the contours in figure~\ref{fig:LCDM_even_all}. Note that the direction of the $w_{\rm DE}-\aMo$ degeneracy changes when $w_{\rm DE}$ passes through $-1$. This is due to the fact that an equation of state significantly smaller than $-1$ can be easily compensated by increasing the pressure of radiation. On the other hand, an equation of state too close to zero would lead to no  cosmic acceleration, which could not be compensated by a very low pressure from radiation.

\item $\aMo-\aBo$ and $w_{\rm DE}-\aBo$ contours: We obtain a first branch with $w_{\rm DE}<-1$ and $\aM<0$, in which case the stability of the perturbations requires $\aB<0$. Indeed, linearizing eq.~\eqref{eq:cs2a} around $\Lambda$CDM and applying eq.~\eqref{Eq:Exp_zeta},
we find that
\begin{equation}
    c^2_s\alpha = 2\aM +3(1+w_{\rm DE})\Omega_{\rm DE}+(1+\Omega_r-3\Omega_{\rm DE})\aB-2\aB'\, . \label{eq:cslin}
\end{equation}
 Since $1+\Omega_r-3\Omega_{\rm DE}<0$ at low redshift, we see that $\aB<0$ is needed to keep $c^2_s\alpha>0$ when the first two terms are negative 
(note that $\aB'$ has the same sign as $\aB$ with our chosen time evolution). Moreover, since $\Omega_{\rm DE}$ and $\aM$ both decay at high redshift, while the term in front of $\aB$ increases at high redshift (it becomes less negative), a strongly negative $\aBo$ is necessary to maintain $c^2_s\alpha>0$ at higher redshift.\footnote{Note that, when dark energy becomes negligible at very high redshift, $1+\Omega_r-3\Omega_{\rm DE}$ changes sign. However, $\aB$ and $\aM$ also approach zero at that point and we recover GR.} This is clearly visible in the $\aMo-\aBo$ and $w_{\rm DE}-\aBo$ contours of figure~\ref{fig:LCDM_even_all}.

On the other hand, we have a second branch with $w_{\rm DE}>-1$ and $\aM>0$, which both contribute positively to the sound speed. Since the term proportional to $\aB$ in eq.~\eqref{eq:cslin} dominates over the first two at high redshift, $\aBo$ needs to be negative also in this branch to keep $c_s^2\alpha>0$ in the past. In this case, however, the absolute value of $\aBo$ does not need to be as large as in the other branch, as is visible from figure~\ref{fig:LCDM_even_all}.

\item Contours involving the parameter $\gamma_{c,0}$: In addition to the distance to last scattering and the stability condition, the parameters are constrained by the even multipoles, which are directly sensitive to the galaxy growth rate $f$. The growth rate depends on the evolution of the dark matter and baryon density given in eqs.~\eqref{Eq:pert_b} and~\eqref{Eq:pert_c}, providing a constraint on the combination 
\begin{equation}
\frac{1}{c_s^2\alpha}(\alpha_{\rm B}-\alpha_{\rm M}+3\gamma_c)(\alpha_{\rm B}-\alpha_{\rm M}+3\gamma_c \omega_c b_c)\,. \label{eq:growth_mod}
\end{equation}
This explains the behavior of $\gamma_{c,0}$ along the aforementioned two branches. Along the first branch, $\aMo$ and $\aBo$ are negative, and $|\aBo|>|\aMo|$ (to ensure stability also in the past). To counterbalance this, and keep the growth rate close to $\Lambda$CDM, it is necessary to have $\gamma_{c,0}>0$. Along the second branch, we have $\aMo>0$ and large, while $\aBo<0$, again requiring $\gamma_{c,0}>0$ to keep the growth rate consistent with $\Lambda$CDM. This is clearly visible in the $\gamma_{c,0}-\aMo$ and $\gamma_{c,0}-\aBo$ contours of figure~\ref{fig:LCDM_even_all}. 

The degeneracy between $\gamma_c$ and $\alpha_{\rm M}$ can be understood from the Einstein and Euler equations: a positive $\gamma_c$ mainly enhances the growth of structure by changing the way dark matter falls in a gravitational potential, as can be seen from eq.~\eqref{Eq:Euler_cdm}. Indeed, the coupling of the scalar field to dark matter generates an additional force on dark matter particles. On the other hand, a positive running of the Planck mass $\alpha_{\rm M}$ in the first branch implies that the mass increases with time, leading to a decrease of the effective Newton constant in the Poisson equation, see eq.~\eqref{Eq:Poisson}. This implies that a given matter density perturbation leads to a smaller distortion of space through the Poisson equation, thus decreasing the growth of structure and counterbalancing the enhancement arising from from $\gamma_c >0$. In the second branch, a negative $\alpha_{\rm B}$ counterbalances a positive $\gamma_c$. The parameter $\alpha_{\rm B}$, which encodes a mixing between the scalar and tensor kinetic terms, impacts the kinetic energy of the scalar field, which in turns affects the Poisson equation and therefore the growth of structure.
\end{itemize}

From figure~\ref{fig:LCDM_even_all}, we notice that the posteriors are highly non-Gaussian, due to the strong degeneracies between the parameters. Such a behavior cannot be captured by a Fisher analysis, which assumes Gaussian posteriors. Therefore, our results show that it is essential to perform an MCMC analysis to fully understand the degeneracies in Horndeski theories and scenarios involving a breaking of the WEP, and correctly forecast the constraining power of the coming generation of surveys.

\subsubsection{Including gravitational redshift by adding the dipole}\label{sec:results_LCDM_dipole}

The inclusion of the dipole in the analysis (red contours in figure~\ref{fig:LCDM_even_all}) has a strong impact on the constraints, favoring one of the branches over the other. This is due to the fact that the dipole is directly sensitive to the combination 
\begin{equation}
\gamma_c \left[f - (1+z)\left(\frac{\mathcal{H}_0}{\mathcal{H}}\right)^2 \, \frac{ 3 \, \Omega_{m,0}}{c_s^2 \alpha}(\aB - \aM + 3 \gamma_c \omega_c b_c)\right]\,,
\end{equation}
as can be seen from eq.~\eqref{Eq:xi1}. From this, we notice that keeping the dipole close to its $\Lambda$CDM value requires either a small $\gamma_c$ or a cancellation of the two terms in the square bracket. This tightens the constraints on $\gamma_c$, as is clearly visible in figure~\ref{fig:LCDM_even_all}. Moreover, the degeneracy between $\gamma_c$, $\aB$ and $\aM$ is broken by imposing that $(\aB-\aM+3\gamma_c w_c b_c)/c_s^2\alpha$ is tuned to balance the growth rate $f$. From figure~\ref{fig:LCDM_even_all}, we see that adding this constraint by including the dipole almost completely removes one of the branches in the parameter space and significantly tightens the other. Since $w_{\rm DE}$ is strongly degenerated with $\aM$, the tightening of the constraints on $\aM$ from the dipole also improves the constraints on $w_{\rm DE}$.

In appendix~\ref{app:plots}, figure~\ref{fig:LCDM_mono_dip}, we show the constraints arising from the monopole alone and compare them to an analysis based on the combination of monopole and dipole. Also in this case, the dipole efficiently breaks the degeneracy between the parameters and removes one of the branches. The only difference with respect to the analysis with all multipoles is that the monopole is much less constraining on its own than in combination with the quadrupole and hexadecapole. This is due to the fact that the growth rate is degenerated with the galaxy bias using the monopole alone, whereas the inclusion of the other two even multipoles breaks this degeneracy and overall reduces the size of the two branches. 

The improvement arising from the dipole is impressive: the constraints on the EFT parameters are tightened by up to 50\%, as can be seen from Table~\ref{tab:LCDMconstraints}. This is especially remarkable considering that the signal-to-noise ratio (SNR) of the dipole is significantly smaller than that of the even multipoles. Indeed, for a bias difference $\Delta b=1$ the cumulative SNR of the dipole in $\Lambda$CDM from separations $d=20-160\,\mbox{Mpc/h}$ over the redshift range $z=0.15-1.55$ is 80, while that of the quadrupole is 461, i.e.\ almost 6 times larger.

The reason why such a small signal can significantly improve the constraints is because deviations from $\Lambda$CDM affect it differently from the even multipoles. Hence, the key element to quantify its constraining power is the amplitude of the deviations with respect to $\Lambda$CDM. For example, if we choose a point at the edge of the 2\,$\sigma$ contours of figure~\ref{fig:LCDM_even_all} with $\alpha_{\rm M,0}=1.3$, $w_{\rm DE}=-0.8$, $\alpha_{\rm B,0}=-0.4$ and $\gamma_{c,0}=0.17$, we find that the SNR of the deviation is 6.2 for the quadrupole. For the dipole, the SNR of the WEP-breaking term (first line in eq.~\eqref{Eq:xi1}) is 3.6, so not even two times smaller. Since this term has a very different dependence on the EFT parameters than the quadrupole, such an SNR is sufficient to significantly tighten the constraints.

The amplitude of the WEP-breaking term in the dipole in eq.~\eqref{Eq:xi1} is directly sensitive to the galaxy bias difference between the two populations of galaxies, which we assume here to be equal to~1. This is what can be typically expected if the sample of galaxies is split into two populations according to their observed flux, for a population of galaxies similar to that of BOSS~\cite{Gaztanaga:2015jrs}. The situation may be different for the galaxies detected by SKA2, for which the bias difference between two magnitude-limited samples may be smaller, decreasing the constraining power of the dipole. In such a case, it would be important to explore alternative ways of splitting galaxies to achieve the breaking of degeneracies we obtain in our forecasts, for example density splits where the galaxies are divided into two groups according to their environment. In this case, the bias difference between the two populations can become significantly larger~\cite{Paillas:2021oli,Beutler:2020evf}. To mimic this, we have boosted the dipole by a factor~2, 5 and 10, see figure~\ref{fig:DBoost} in appendix~\ref{app:plots}. We find that the contours are significantly tightened when the dipole is boosted, completely removing the second branch. This motivates the search for estimators of the dipole that would enhance the gravitational redshift contribution. 

Finally, we notice that, while the dipole tightens the constraints on the model parameters, we do not see a significant impact on the nuisance parameters $X$, $b_{1,\rm B}$, $b_{2,\rm B}$, $b_{1,\rm F}$ and $b_{2,\rm F}$. Concerning the four parameters specifying the biases of the two galaxy populations, this result matches our expectation that galaxy bias can be well measured from the even multipoles alone. The constraints on the relation $X$ between the background CDM fraction and the average fraction of CDM inside a galaxy are always completely dominated by the physical prior specified below eq.~\eqref{eq:xc_xb}, which luckily does not spoil the constraints on the model parameters. The corner plot for the full parameter space is included in appendix~\ref{app:plots}, figure~\ref{fig:LCDMall}, showing the constraints from the even multipoles alone and from all multipoles. 

We could further extend our analysis by varying the cosmological parameters $\Omega_m$, $\Omega_b$, $\Omega_r$ at recombination as well as $A_s$ and $n_s$ in a joint analysis with CMB data. 
This would slightly broaden the constraints on our set of parameters. However, we do not expect this to affect the message of our work. A priori, letting the cosmological parameters vary should affect in a similar way the even multipoles and the dipole, such that the percentage improvement brought by the dipole should remain roughly the same.

\subsubsection{Changing the parameterization for the dark energy evolution}
\label{sec:DEparam1}

\begin{figure*}
    \centering    
 \includegraphics[width=.75\textwidth]{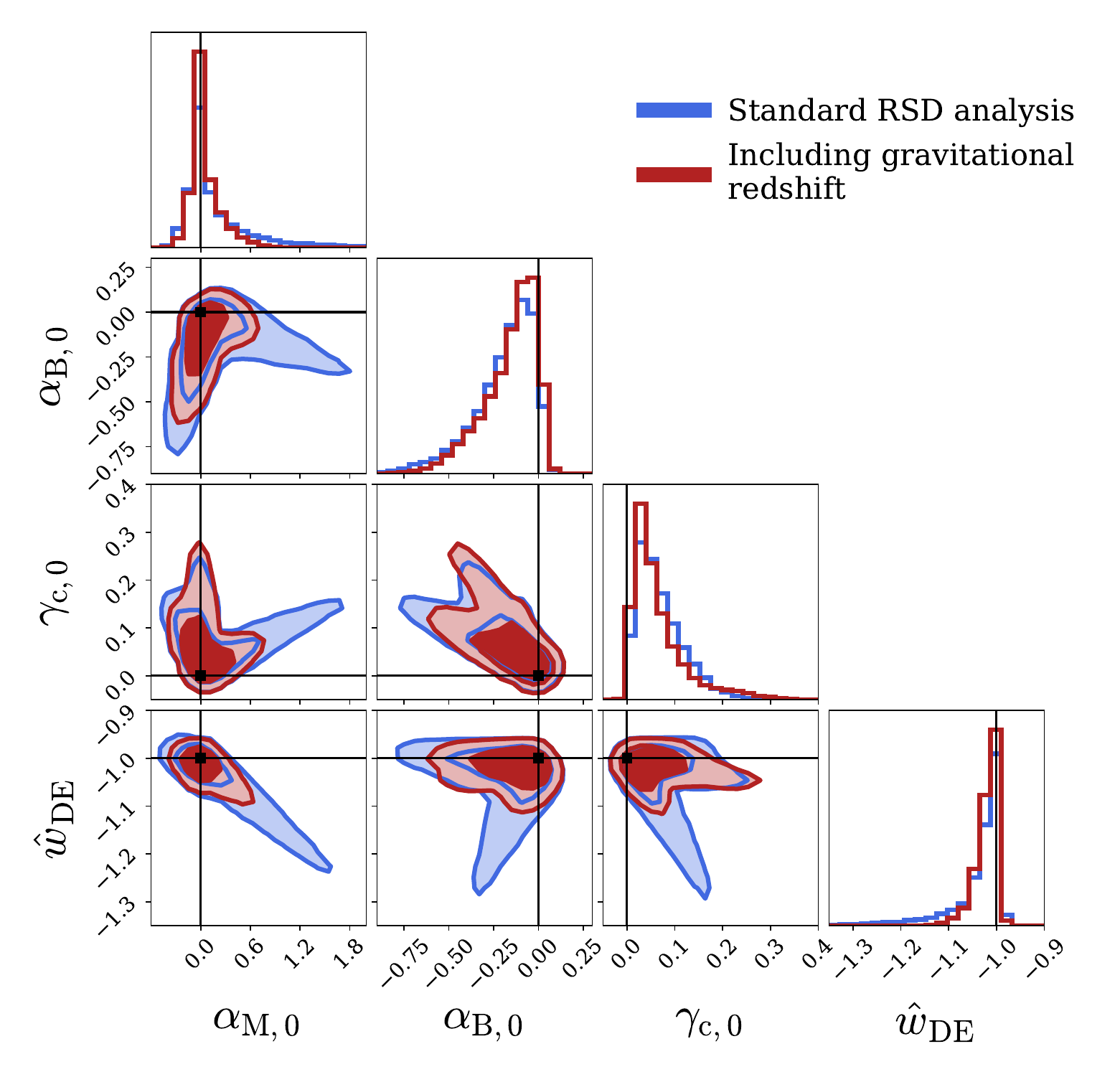}
    \caption{Constraints around a $\Lambda$CDM fiducial model, employing the alternative parameterization with the effective equation of state parameter $\hat{w}_{\rm DE}$.} \label{fig:LCDM_even_allOLD}
\end{figure*}

\begin{table}
\renewcommand{\arraystretch}{1.5}
  \centering
  \begin{tabular}{l|rr|ccr}
    \hline\hline
     & \multicolumn{2}{c|}{Median}& \multicolumn{3}{c}{Width of the $68\%$ credible interval}  \\
      &No dipole & Dipole & No dipole & Dipole & Reduction  \\
    \hline\hline
    $\alpha_{\rm M,0}$ & $0.02$ & $0.00$  & $[-0.12, 0.47]$ & $[-0.08, 0.20]$ & 53\% \\
    $\alpha_{\rm B,0}$ & $-0.15$ & $-0.12$ & $[-0.20, 0.12]$ & $[-0.19, 0.10]$ & 9\% \\
    $\gamma_{\rm c,0}$ & $0.06$ & $0.05$ & $[-0.03, 0.06]$ & $[-0.03, 0.06]$ & 0\%  \\
    ${\hat w}_{\rm DE}$ & $-1.02$ & $-1.01$& $[-0.06, 0.02]$ & $[-0.03, 0.01]$ & 50\% \\
    \hline\hline
  \end{tabular} 
  \caption{Median values and 68\% credible intervals for the $\Lambda$CDM fiducial model both for the standard RSD analysis (no dipole) and when adding the dipole that contains gravitational redshift, employing the alternative parameterization with the effective equation of state parameter $\hat{w}_{\rm DE}$. For the credible interval, we also show the percent reduction obtained by the latter. \label{tab:LCDMconstraints_altparameterization}}
\end{table}

The equation of state parameter of dark energy, $w_{\rm DE}$, is defined as the ratio of the dark energy pressure and the dark energy density. From eq.~\eqref{eq:evol_rho_DE}, we see that the evolution of the dark energy density depends on $w_{\rm DE}$ and also on $\aM$ and $\gamma_c$, where the latter governs the interaction of dark energy with dark matter. Alternatively, we can define an effective equation of state, $\hat{w}_{\rm DE}$, directly encoding the evolution of dark energy through 
\begin{align}\label{eq:time_evol_DE_eff}
\bar \rho_{\rm DE}'=-3(1+\hat{w}_{\rm DE})\bar \rho_{\rm DE} \, .
\end{align}
Comparing eq.~\eqref{eq:time_evol_DE_eff} with eq.~\eqref{eq:evol_rho_DE}, we see that the effective equation of state is related to the true equation of state by
\begin{align}\label{eq:EOS_DE_eff}
\hat{w}_{\rm DE}= w_{\rm DE} -\frac{1}{3}\aM\frac{\bar \rho_{\rm tot}}{\bar\rho_{\rm DE}}+\gamma_c\frac{\bar\rho_c}{\bar\rho_{\rm DE}}\, .
\end{align}
This leads to the following expression for the Hubble parameter
\begin{align}
H(z)=H_0\exp\left[\frac{3}{2}\int_{\ln(a(z))}^0 \hspace{-0.5cm}\mathrm d\ln a'\left(1+\frac{1}{3}\alpha_{\rm M}+\frac{1}{3}\Omega_r+\hat{w}_{\rm DE}\Omega_{\rm DE} -\gamma_{c} \Omega_{c}\right) \right]\, . \label{eq:H_backward_param1}
\end{align}
In figure~\ref{fig:LCDM_even_allOLD}, we show the constraints on the new set of parameters $\alpha_{\rm M,0}$, $\alpha_{\rm B,0}$, $\gamma_{c, 0}$ and $\hat{w}_{\rm DE}$. As in the previous analysis, we see that the inclusion of the dipole significantly alleviates the strong degeneracies among the parameters that would otherwise be present. The direction of the degeneracies is however different with the new parameterization, which is not surprising given that the relation between $w_{\rm DE}$ and $\hat{w}_{\rm DE}$ depends on $\gamma_c$ and $\aM$. In particular, we notice that now an equation of state $\hat{w}_{\rm DE}<-1$ is degenerated with a positive $\aM$. This follows from the angular diameter distance constraint, which depends on the evolution of $H(z)$. In this new parameterization, the latter depends directly on $\aM$, $\gamma_c$ and $\hat{w}_{\rm DE}$, as can be seen from eq.~\eqref{eq:H_backward_param1}. From this equation we notice that $\hat{w}_{\rm DE}<-1$ is counterbalanced by $\aM>0$ and $\hat{w}_{\rm DE}>-1$ by $\aM<0$. As before, this leads to two branches in the parameter space when only the even multipoles are considered.

We see that the constraints are clearly asymmetric around $\Lambda$CDM. This is particularly visible in the $\alpha_{\rm M,0}-\hat{w}_{\rm DE}$ contours, which are shifted towards positive values of $\alpha_{\rm M,0}$. This asymmetry is related to an interplay between the stability conditions and the constraints from the growth rate and the angular diameter distance. Indeed, linearizing the combination $c^2_s \alpha$ around $\Lambda$CDM with the new parameterization, we find
\begin{equation}
    c^2_s\alpha = 3\aM+3(1+\hat{w}_{\rm DE})\Omega_{\rm DE} -(2-3\Omega_m-4\Omega_r)\aB-2\aB'-3\gamma_c\Omega_c\, . \label{eq:cslin_old}
\end{equation}
The degeneracy in the $\alpha_{\rm M,0}-\hat{w}_{\rm DE}$ plane is governed by the distance to last scattering, which is sensitive to the combination $\aM+3(1+\hat{w}_{\rm DE})\Omega_{\rm DE}$ through eq.~\eqref{eq:H_backward_param1}. Along this degeneracy direction, when $\aM$ is positive, the positivity of the combination $\aM+(1+\hat{w}_{\rm DE})\Omega_{\rm DE}$ in eq.~\eqref{eq:cslin_old} is automatically maintained. This leads to a positive $c^2_s \alpha$ even when $\aB=0$, as long as $\gamma_c$ is small enough. Hence, the stability condition and the requirement of having a distance and growth rate close to $\Lambda$CDM are simultaneously satisfied. On the other hand, when $\aM$ is negative along the degeneracy direction, then $\aM+(1+\hat{w}_{\rm DE})\Omega_{\rm DE}<0$. To ensure stability, it is thus necessary to have negative values of $\aB$ and $\gamma_c$. This leads to stronger deviations in the growth rate, which are disfavored. 

As with the other parameterization, we see that adding the dipole strongly tightens the constraints on $\alpha_{\rm M,0}$ by breaking the degeneracies with $\alpha_{\rm B, 0}$ and $\gamma_{c,0}$, which in turns also tightens the constraints on $\hat{w}_{\rm DE}$. Quantitatively, from Table~\ref{tab:LCDMconstraints_altparameterization}, we see that the 1\,$\sigma$ constraints on $\alpha_{\rm M,0}$ and $\hat{w}_{\rm DE}$ are improved by 53\% and 50\% respectively when the dipole is added. On the other hand, the constraints on $\aBo$ are only mildly improved with this parameterization, while those on $\gamma_{c,0}$ show no improvement. Nevertheless, the red 2-dimensional contours involving $\gamma_{c,0}$ in figure~\ref{fig:LCDM_even_allOLD} show a clear preference of one branch when including gravitational redshift. Overall, comparing Table~\ref{tab:LCDMconstraints} and~\ref{tab:LCDMconstraints_altparameterization}, we see that the dipole significantly tightens the constraints in both cases, but the way the improvement is distributed among the parameters clearly depends on the parameterization.

\subsection{Constraints around modified gravity fiducials}
\label{sec:MGres}

\begin{figure*}
    \centering
    \includegraphics[width=.75\textwidth]{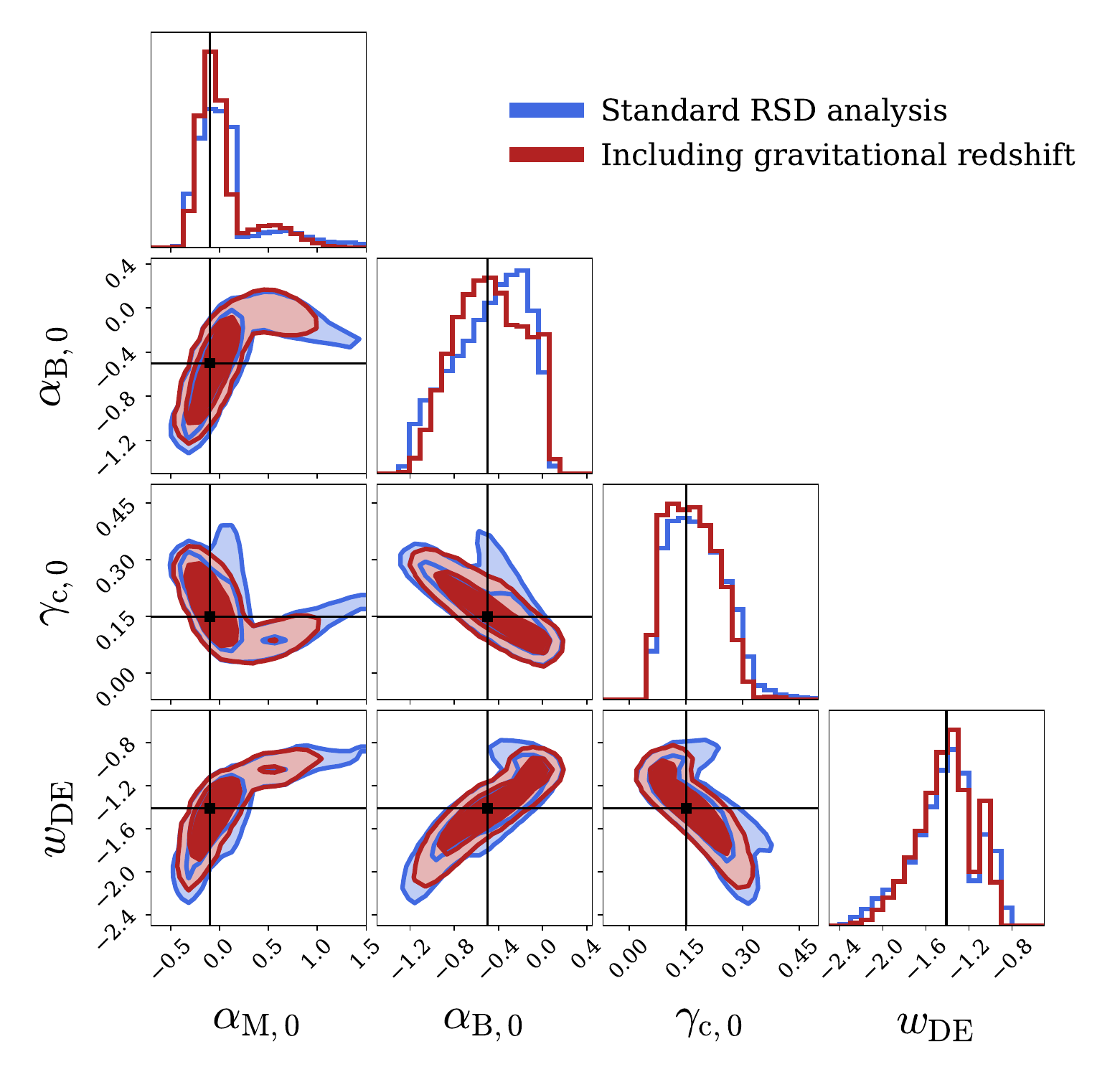}
    \caption{Constraints around the modified gravity fiducial MGI.}\label{fig:MG}
\end{figure*}

As explained in section~\ref{sec:fiducials}, we also explore constraints around two other fiducial models: one that would solve the Hubble tension (MGI) and another one that is closer to $\Lambda$CDM (MGII). The results for MGI are plotted in figure~\ref{fig:MG}. We notice that the double branches around MGI are less pronounced in the blue contours (standard RSD analysis), than around $\Lambda$CDM (figure~\ref{fig:LCDM_even_all}). This is due to the fact that the MGI fiducial model is well inside one of the branches of the $\Lambda$CDM contours. For example, we can see that the point with $\gamma_{c,0}=0.15$ and $\alpha_{\rm M, 0}=-0.1$ are well within the vertical branch of the $\gamma_{c,0}-\alpha_{\rm M, 0}$ contour in figure~\ref{fig:LCDM_even_all}. As such, when we shift the fiducial model to this region, the degeneracies are already partially broken. While of course new degeneracies can appear around this new fiducial, figure~\ref{fig:MG} indicates that they are much less pronounced. As a consequence, including gravitational redshift still improves the constraints, but in a less significant way than around $\Lambda$CDM. From figure~\ref{fig:MG}, we also notice that if MGI were to be the true model in our Universe, data from SKA2 would not be able to rule out $\Lambda$CDM based on the gravity + scalar sector only, as the 2\,$\sigma$ error bars on $\alpha_{\rm M,0}$, $\alpha_{\rm B,0}$ and $w_{\rm DE}$ encompass the corresponding $\Lambda$CDM values. However, we see that SKA2 would be able to favor MGI over $\Lambda$CDM based on the coupling to dark matter $\gamma_{c,0}$, which is incompatible with zero at more than 2\,$\sigma$.  Note that a model different from MGI might lead to a different conclusion regarding the relation with $\Lambda$CDM. 

The constraints around MGII are shown in appendix~\ref{app:plots}, figure~\ref{fig:MGsmall}. In this case, the degeneracies are still very pronounced with a standard RSD analysis. Since MGII is only mildly away from $\Lambda$CDM, the two branches around this point are quite similar to those around $\Lambda$CDM. As a consequence, we see that adding gravitational redshift significantly tightens the constraints, in a similar way as for the $\Lambda$CDM fiducial.

\subsection{Constraints for specific sub-classes of Horndeski theories}
\label{sec:LCDMfix}

\begin{figure*}
    \centering
    \includegraphics[width=.495\textwidth]{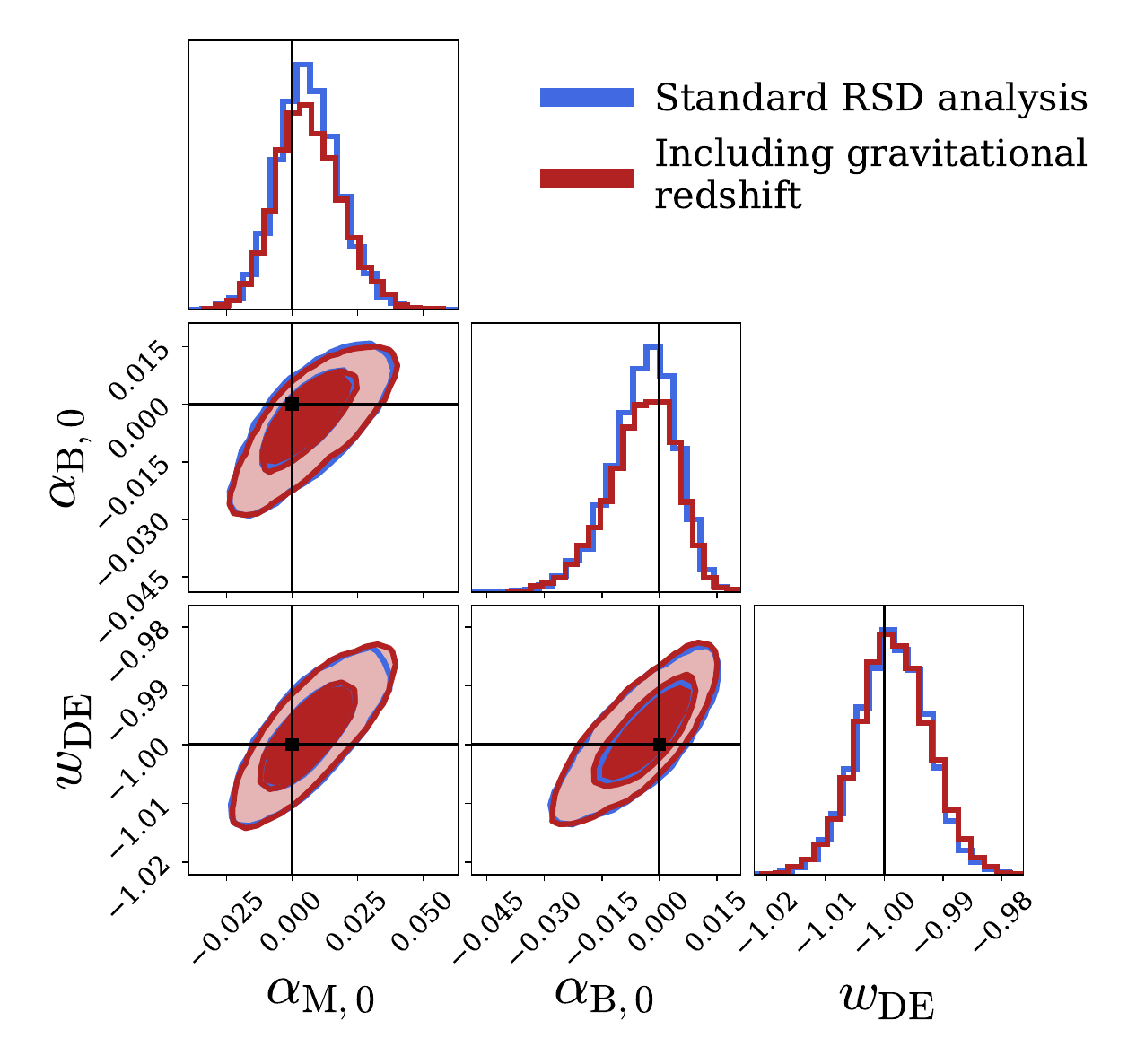}
    \includegraphics[width=.495\textwidth]{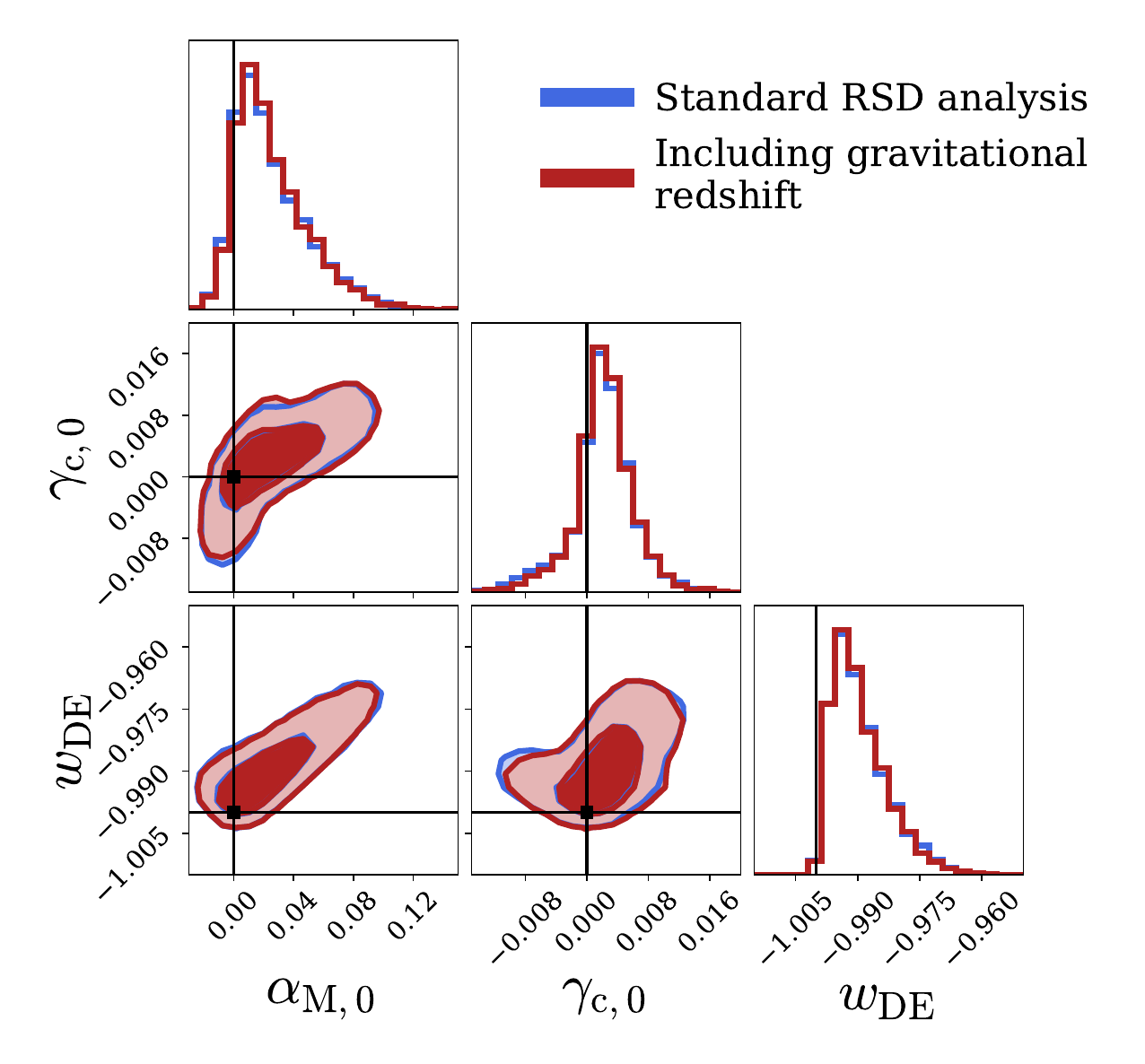}
    \caption{Constraints given a $\Lambda$CDM fiducial when one parameter is kept fixed. \emph{Left panel}: $\gamma_c$ is fixed to zero. \emph{Right panel}: $\alpha_{\rm B} = \alpha_{\rm M}/2$. }\label{fig:LCDM_fix}
\end{figure*}

As discussed above, the strechangeh of gravitational redshift is to break the degeneracy between the parameters $\gamma_{c,0}$, $\aBo$ and $\aMo$. As a comparison, we investigate specific models where only two of the three parameters are left free. In figure~\ref{fig:LCDM_fix} (left panel), we show the contours for the case of Horndeski theories without a breaking of the WEP, i.e.~with $\gamma_c=0$. In the right panel we show another specific case, where the parameter $\gamma_{c,0}$ is left free, but we fix $\aB=\aM/2$. This corresponds to various Horndeski theories (cf.\ Table~1 in ref.~\cite{Bellini:2014fua}), including Brans-Dicke theories~\cite{Brans:1961sx, Sawicki:2012re} and some $f(R)$ models~\cite{Carroll:2003wy, Song:2006ej, Carroll:2006jn, Vollick:2003aw}. From these figures, we see that the constraints from the RSD analysis are much tighter than before. The degeneracies in the parameter space are indeed much less severe when one parameter is removed. In particular, the strong degeneracy between $\aMo$ and $w_{\rm DE}$ is reduced, due to the increased constraints on $\aMo$ from the growth of structure $f$. 

In these cases, gravitational redshift does not improve the constraints further. This is consistent with the SNR comparison made above: the statistical power of the dipole term (the full expression in eq.~\eqref{Eq:xi1}) is $\sim 6$ times smaller than that of the quadrupole. As such, when there is no strong degeneracy to break, the dipole has almost no impact on the constraints. This is also in agreement with the results of ref.~\cite{Lorenz:2017iez}, who showed that relativistic effects do not help to constrain Horndeski theories with no breaking of the WEP. 

A word of caution is however necessary concerning these results: both the analysis of ref.~\cite{Lorenz:2017iez} and our analysis assume a fixed time evolution for the parameters. Even if we have chosen this time evolution to be the same for all parameters in order to minimize its impact on the degeneracies, the simple fact that we assume a known time evolution helps breaking some of the degeneracies. For example, in eq.~\eqref{eq:growth_mod}, the term $w_c b_c$ has a different time evolution from $\gamma_c$ and $\aM$ and therefore helps breaking the degeneracy between these parameters in the model with $\aB=\aM/2$, when we combine multiple redshift bins. If we consider one redshift bin only, i.e.\ we do not exploit the fact that we have a fixed time evolution for the parameters, we find that the parameters are completely degenerated and cannot be constrained anymore.

\section{Conclusions} \label{sec:conclusion}

In this paper, we show that the effect of gravitational redshift, which will be measurable with the coming generation of galaxy surveys, is crucial to efficiently test models beyond GR. To determine the constraining power of this new observable in a generic way, we have considered the effective theory of interacting dark energy, which encompasses all scalar-tensor theories of gravity and allows for a non-minimal coupling of dark matter. We find that adding gravitational redshift to a standard RSD analysis significantly improves the constraints on the EFT parameters, by up to 50 percent. The precise results depend on the choice of parameterization within the EFT framework, on the time evolution assumed for the EFT functions and on the fiducial cosmological model, but the overall message remains unaffected by these details: gravitational redshift breaks degeneracies between parameters, leading to a significant improvement in the constraints.

To measure this novel observable, no new data are involved: one can simply fit for a dipole in the cross-correlation function between two different populations of galaxies. This requires minimal changes to the analysis pipeline, but leads to a strong impact on the constraints. The improvement is especially remarkable given that the SNR of the dipole is typically six times smaller than that of the standard quadrupole term. However, the specificity of the dipole is that it contains a particular combination of gravitational redshift and Doppler effects, which exactly vanishes if the WEP is valid. This combination is not present in the standard even multipoles of the correlation function and is able to break the degeneracies between the EFT parameters. Note that, since this combination is proportional to the bias difference between the two populations of galaxies, we expect the breaking of degeneracies to scale with the value of this difference. Here, we have assumed a bias difference equal to 1, but estimators that would boost this difference may improve the constraints further, while a lower difference may result in an overall worsening. 

In this work we have focused on spectroscopic surveys, which are directly sensitive to the growth rate of structure through RSD. In a future work, we will study whether the inclusion of gravitational lensing can further improve the constraints. Since gravitational lensing is sensitive to the sum of the two gravitational potentials (the spatial and temporal perturbations in the metric), while gravitational redshift only depends on the distortion of time, combining the two may lead to more stringent constraints on the effective theory of interacting dark energy. Possible future work could also include an investigation beyond the quasi-static approximation assumed in our work and the choice of other time parameterizations for the EFT parameters.

We conclude that measuring gravitational redshift provides an efficient way of increasing the constraining power of large-scale structure surveys at minimal cost, and a combination with additional observables could further increase its remarkable constraining power. 

\section*{Software availability}
The Python code \texttt{EF-TIGRE} (\textit{Effective Field Theory of Interacting dark energy with Gravitational REdshift}) developed for this project is publically available on Github at \\\href{https://github.com/Mik3M4n/EF-TIGRE}{github.com/Mik3M4n/EF-TIGRE}. The version used for this paper is available on Zenodo at~\href{https://zenodo.org/doi/10.5281/zenodo.10606418}{10.5281/zenodo.10606418}.

\acknowledgments
We thank Lawrence Dam for useful discussions and Goran Jelic-Cizmek for helpful comments concerning the numerical implementation of the covariance. The numerical analysis was performed on the Baobab computer cluster of the University of Geneva. S.C., N.G., I.T. and C.B. acknowledge support from the European Research Council (ERC) under the European Union’s Horizon 2020 research and innovation program (grant agreement No. 863929; project title “Testing the law of gravity with novel large-scale structure observables”). D.S.B. and C.B. acknowledge support from the Swiss National Science Foundation. The work of M.M. is supported by European Union's H2020 ERC Starting Grant No.~945155--GWmining, Cariplo Foundation Grant No.~2021-0555, the ICSC National Research Centre funded by NextGenerationEU, and MIUR PRIN Grant No. 2022-Z9X4XS.

\appendix

\section{Relation to the phenomenological parameters} \label{app:phenomenological}

The effective theory of interacting dark energy adopted in this work only contains a limited number of free functions ($\aM$, $\aB$, $\gamma_c$ and $w_{\rm{DE}}$), which all have a simple theoretical interpretation and enter at the level of the Lagrangian. However, another common approach to investigate the impact of modified gravity and dark energy on cosmological observables consists in introducing some phenomenologically-motivated free functions, entering directly in the Poisson and Euler equations (see e.g.\ refs.~\cite{Amendola:2007rr,Bertschinger:2008zb,Silvestri:2013ne,Song:2010fg,Pogosian:2010tj,Pogosian:2016pwr,Amendola:2019laa}). In this appendix, we state the relation between the two sets of parameters arising from these different approaches.

To parameterize modifications in the Poisson equation, the gravitational constant $G$ is usually multiplied by a free function $\mu$, as done in eq.~\eqref{Eq:Poisson}. Indeed, $\mu$ can be expressed using the free functions $\aM$, $\aB$ and $\gamma_c$, see eq.~\eqref{Eq:mu}. In the phenomenological approach, the difference between the two potentials $\Phi$ and $\Psi$ is expressed by the gravitational slip $\eta$,
\begin{equation}
    \eta =  \Phi/ \Psi\,.
\end{equation}
The gravitational slip is obtained by combining eqs.~(4.26) and (4.27) of ref.~\cite{Gleyzes:2015pma} (note that their $\Phi$ and $\Psi$ are interchanged compared to our notation), and translates into
\begin{equation}
    \eta = \frac{c_s^2\alpha+2\aB(\aB-\aM+{3\gamma_c\omega_cb_c})}{c_s^2\alpha+2(\aB-\aM){(\aB-\aM+3\gamma_c\omega_cb_c)}}\,. \label{eq:eta}
\end{equation}

Apart from the phenomenologically motivated functions $\mu$ and $\eta$, ref.~\cite{Bonvin:2018ckp} has introduced the free functions $\Theta$ and $\Gamma$ encompassing a breaking of the WEP for CDM. They appear as a friction and fifth force term in the Euler equation,
\begin{equation}
    V'_{\rm c}+(1+\Theta)V_{\rm c}-\frac{k}{\mathcal{H}}(1+\Gamma)\Psi = 0\,. \label{Eq:BreakingWEP}
\end{equation}
Comparing this with the Euler equation of the effective theory of interacting dark energy given in eq.~\eqref{Eq:Euler_cdm}, we obtain
\begin{equation}
    \Theta = 3\gamma_c\,, \label{Eq:Theta}
\end{equation}
and
\begin{equation}
    \Gamma 
    =\frac{6\gamma_c \rbr{\aB - \aM + 3 \gamma_c \omega_c b_c}}{c_s^2 \alpha + 2 (\aB - \aM) \, (\aB - \aM + 3 \gamma_c \omega_c b_c)}\,. \label{Eq:Gamma}
\end{equation}

Note that these phenomenological modifications enter the perturbation equations as follows (see ref.~\cite{Castello:2022uuu}),
\begin{align}
    & \delta_b''+ (2 + \zeta) \delta_b' - \frac{3}{2} \Omega_m \delta_m \, \mu = 0\,, \\
    & \delta_c''+ (2 + \zeta + 3 \gamma_c) \delta_c' - \frac{3}{2} \Omega_m \delta_m \, \mu(\Gamma + 1) = 0\,,
\end{align}
which, combined with eqs.~\eqref{Eq:mu}, \eqref{Eq:Theta} and \eqref{Eq:Gamma}, indeed recover eqs.~\eqref{Eq:pert_b} and~\eqref{Eq:pert_c}.

Galaxy surveys have provided constraints on the free functions $\mu$ and $\eta$ through gravitational lensing~\cite{DES:2018ufa, Planck:2018vyg,DES:2022ccp} and RSD measurements~\cite{eBOSS:2020yzd}. The combination of data from the CMB, baryon acoustic oscillations, supernovae, weak lensing and RSD recently allowed for a reconstruction of their redshift evolution~\cite{Pogosian:2021mcs} as well as their joint redshift and scale dependence~\cite{Garcia-Quintero:2020bac}. Ref.~\cite{Castello:2022uuu} pointed out that the standard RSD-based constraints on $\mu$ are strongly degenerate with the parameters $\Theta$ and $\Gamma$ describing a breaking of the WEP, and forecasted constraints when taking gravitational redshift into account.

We emphasize that, despite the existence of analytical expressions relating the free functions of the phenomenological and effective theory framework, these two approaches are fundamentally different. First of all, in the phenomenological approach, the free functions $\mu$, $\eta$, $\Gamma$ and $\Theta$ are usually assumed to be independent. However, we see in eqs.~\eqref{Eq:mu}, \eqref{eq:eta}, \eqref{Eq:Theta} and \eqref{Eq:Gamma} that they in fact display dependencies through more fundamental, theory-based free functions. Moreover, while we assume that the theory-based functions evolve according to the dark energy density, see eq.~\eqref{eq:p_evolution}, their interplay in the equations for the phenomenologically motivated functions leads to a different time dependence. Hence, a time evolution of $\mu$, $\eta$, $\Gamma$ and $\Theta$ proportional to the dark-energy density (as applied e.g.~in refs.~\cite{DES:2018ufa, Planck:2018vyg, eBOSS:2020yzd, Castello:2022uuu}) cannot be maintained in our effective theory framework. 

To summarize, a direct, quantitative comparison between the constraints in our work to those obtained employing the phenomenologically motivated parameters cannot be performed. However, some qualitative results are reflected in both these approaches. Most importantly, the role of gravitational redshift in breaking degeneracies that was pointed out in ref.~\cite{Castello:2022uuu} is also maintained in our effective theory approach, as extensively discussed in section~\ref{sec:results}.

\section{Supplementary plots}
\label{app:plots}

\begin{figure*}[ht]
    \centering
    \includegraphics[width=.85\textwidth]{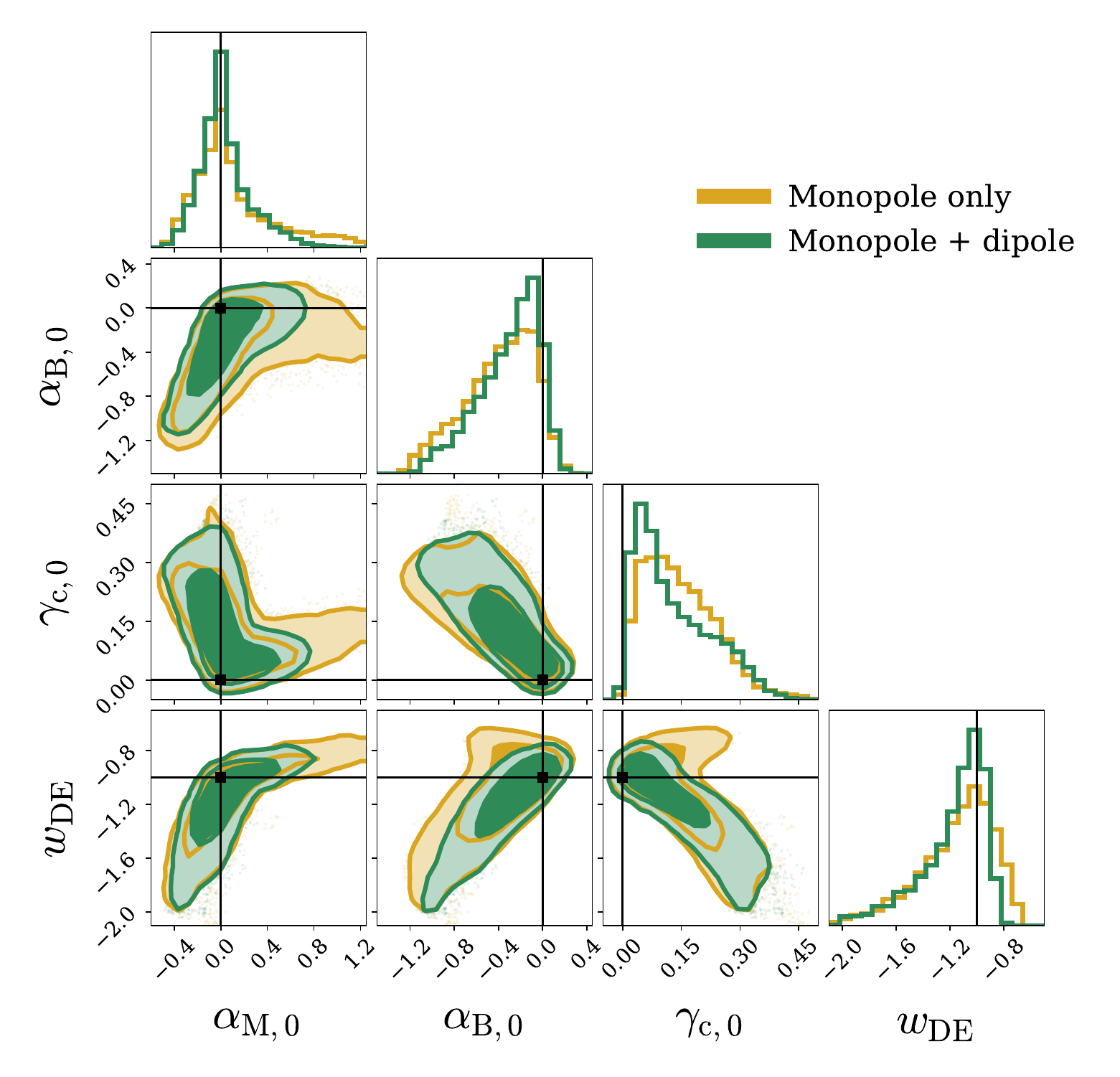}
    \caption{Constraints around the $\Lambda$CDM fiducial when including only the monopole, and combining monopole and dipole.} \label{fig:LCDM_mono_dip}
\end{figure*}

\begin{figure*}[ht]
    \centering 
    \includegraphics[width=.85\textwidth]{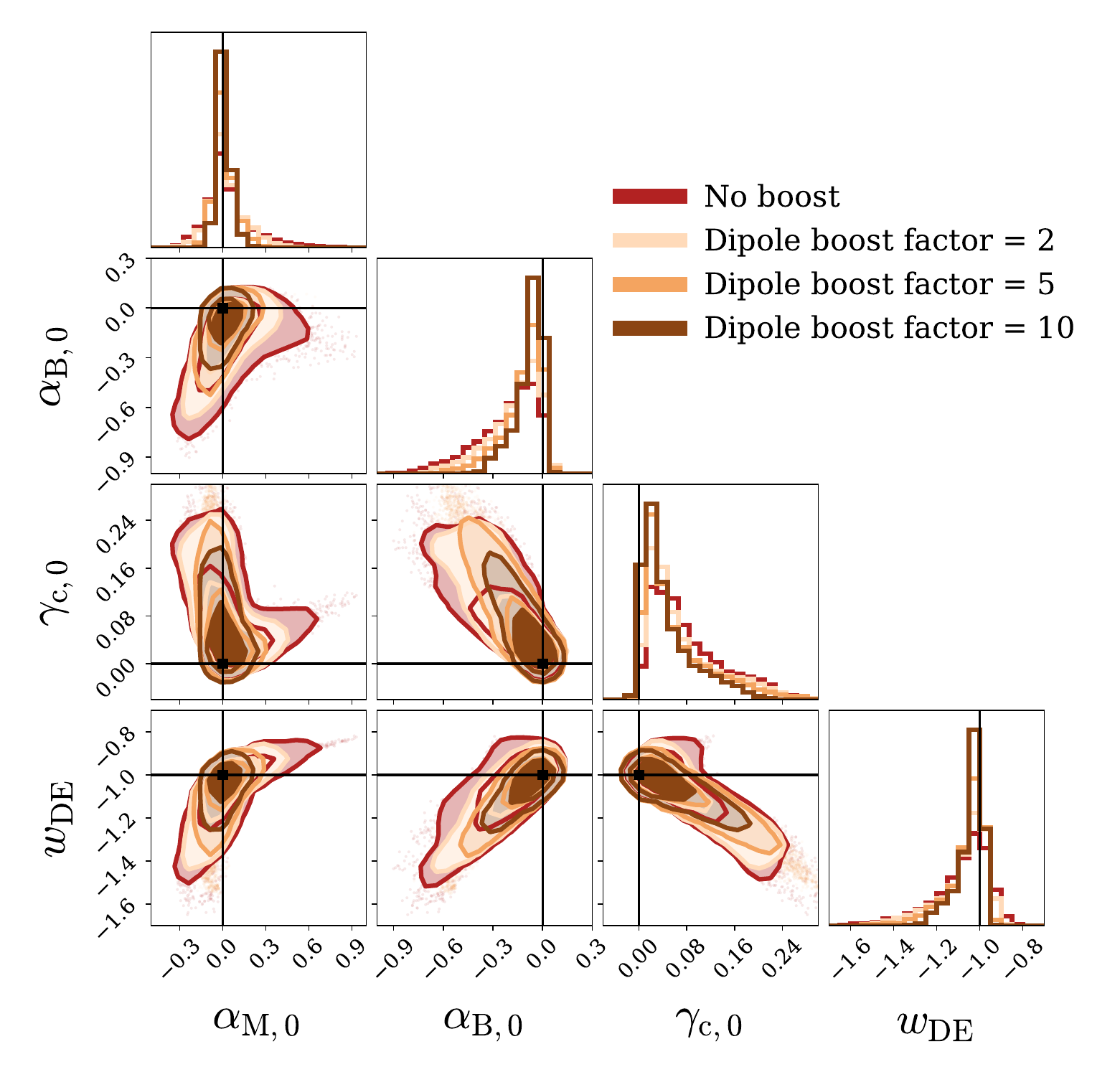}
    \caption{Constraints around the $\Lambda$CDM fiducial when including all multipoles and applying various boost factors to the dipole.}\label{fig:DBoost}
\end{figure*}

\begin{figure*}
    \centering
    \includegraphics[width=\textwidth]{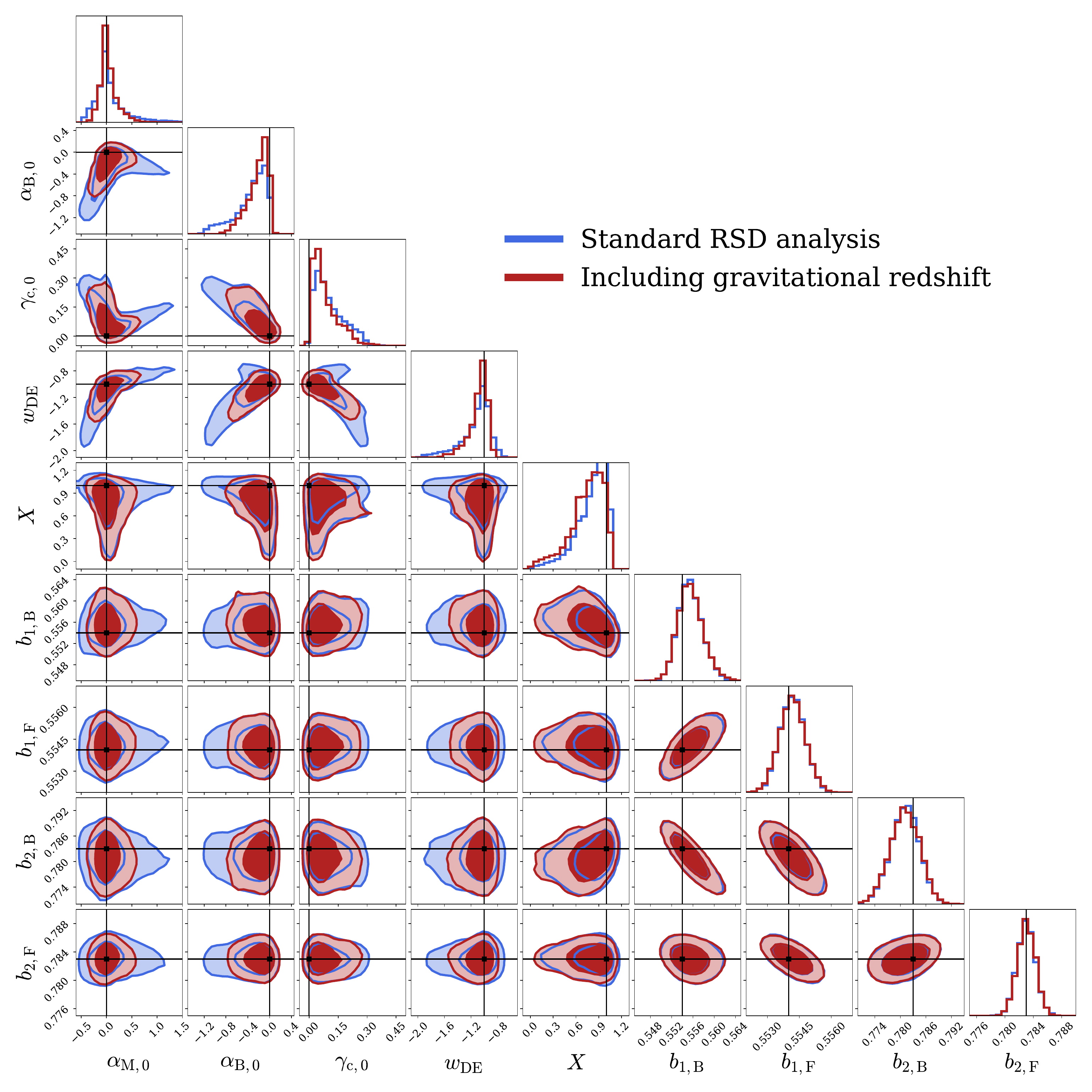}
    \caption{Constraints around a $\Lambda$CDM fiducial for the full parameter space.} \label{fig:LCDMall}
\end{figure*}

\begin{figure*}
    \centering \includegraphics[width=.85\textwidth]{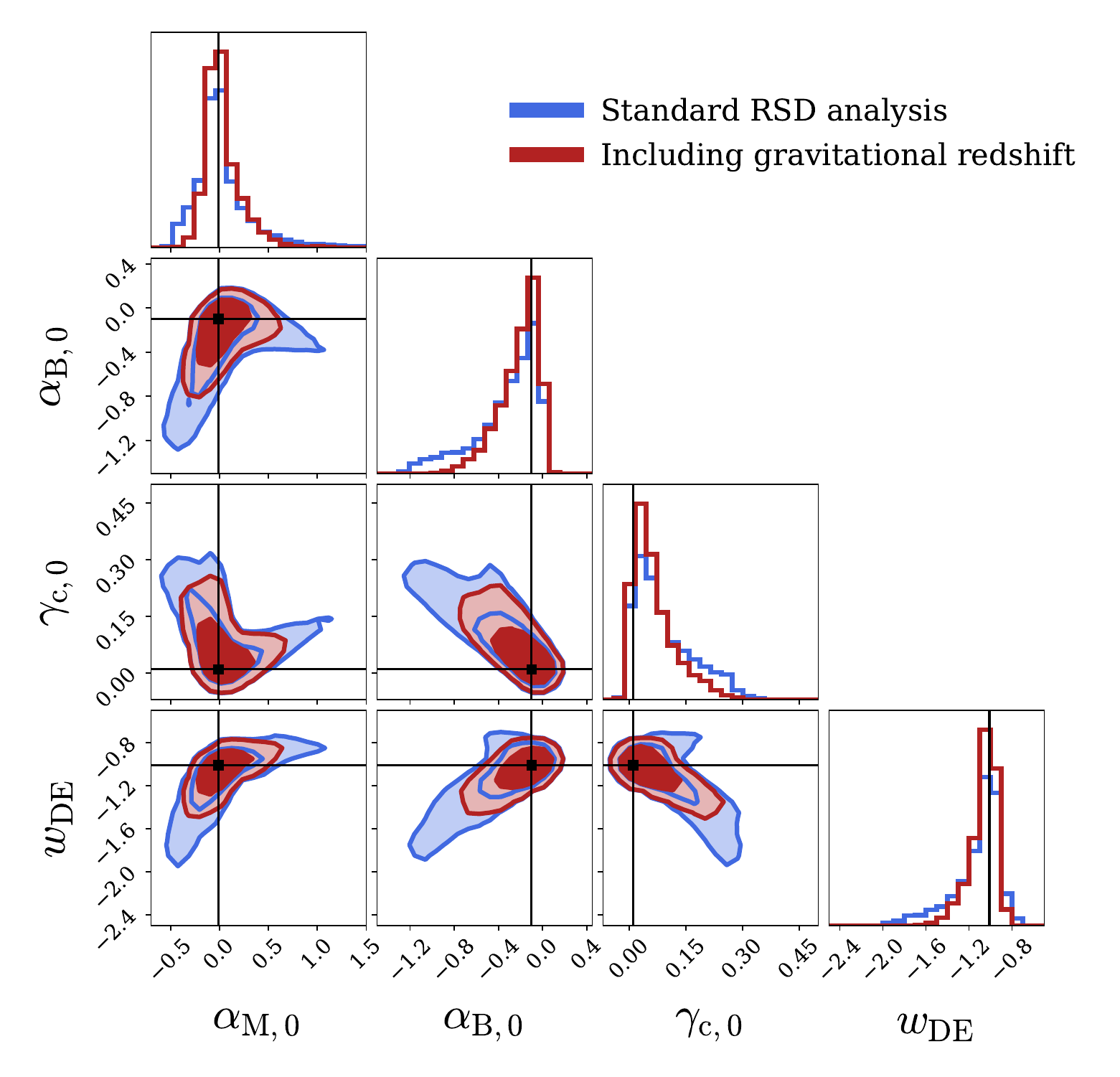}
    \caption{Constraints around the modified gravity fiducial MGII. The constraints are very similar to those around $\Lambda$CDM, see figure~\ref{fig:LCDM_even_all} }\label{fig:MGsmall}
\end{figure*}

Here, we present additional plots, supplementing those presented in section~\ref{sec:results}. All results in this appendix, except figure~\ref{fig:MGsmall}, are obtained using the $\Lambda$CDM fiducial model. First, the corner plot in figure~\ref{fig:LCDM_mono_dip} shows the constraints obtained on the parameters $\aMo$, $\aBo$, $\gamma_{c,0}$ and $w_{\rm DE}$ when including only the monopole of the galaxy correlation function (yellow contours), compared to the combination with the dipole term, which contains gravitational redshift (green contours). In figure~\ref{fig:DBoost}, we further illustrate the impact of a dipole boosted by factors 2, 5, and 10 (beige--brown shaded contours), compared to the base case (red contours) without a boost. In figure~\ref{fig:LCDMall} we show the full corner plot, including all parameters specified in eq.~\eqref{Eq:params}, for the standard RSD analysis (blue contours) and also including gravitational redshift (red contours). Finally, in figure~\ref{fig:MGsmall} we show the constraints around the modified gravity fiducial model MGII. The contours are very similar to those obtained around $\Lambda$CDM, see figure~\ref{fig:LCDM_even_all}.

\FloatBarrier

\bibliographystyle{JHEP}
\bibliography{IDE_dipole}

\end{document}

%% file: macros_paper.tex
\newcommand\ee{\end{equation}}
\newcommand\be{\begin{equation}}
\newcommand\eea{\end{eqnarray}}
\newcommand\bea{\begin{eqnarray}}

\newcommand{\B}{\textrm{B}}
\newcommand{\F}{\textrm{F}}

\newcommand{\HH}{\mathcal{H}}

\newcommand{\aT}{\alpha_{\rm{T}}}
\newcommand{\aK}{\alpha_{\rm{K}}}
\newcommand{\aB}{\alpha_{\rm{B}}}
\newcommand{\aM}{\alpha_{\rm{M}}}
\newcommand{\aBo}{\alpha_{\rm{B},0}}
\newcommand{\aMo}{\alpha_{\rm{M},0}}

\newcommand{\rbr}[1]{\left(#1\right)}

